\documentclass[aps,twocolumn,prb,showpacs,floatfix]{revtex4}

\usepackage{epsfig,amsmath,amssymb}
\usepackage{graphics} 
\usepackage{subfigure}
\usepackage{color}
\usepackage{multirow}

\begin{document}

\title
{
A frustrated spin-1/2 Heisenberg antiferromagnet on a chevron-square lattice
}
\author
{P.~H.~Y.~Li and R.~F.~Bishop}
\affiliation
{School of Physics and Astronomy, Schuster Building, The University of Manchester, Manchester, M13 9PL, UK}

\author
{C.~E.~Campbell}
\affiliation
{School of Physics and Astronomy, University of Minnesota, 116 Church Street SE, Minneapolis, Minnesota 55455, USA}

\begin{abstract}
  The coupled cluster method (CCM) is used to study the
  zero-temperature properties of a frustrated spin-half
  ($s=\frac{1}{2}$) $J_{1}$--$J_{2}$ Heisenberg antiferromagnet (HAF)
  on a two-dimensional (2D) chevron-square lattice.  On an underlying
  square lattice each site of the model has 4 nearest-neighbor
  exchange bonds of strength $J_{1}>0$ and 2 frustrating
  next-nearest-neighbor (diagonal) bonds of strength $J_{2} \equiv
  \kappa J_{1}>0$, such that each fundamental square plaquette has
  only one diagonal bond.  The diagonal $J_{2}$ bonds are arranged in
  a chevron pattern such that along one of the two basic square axis
  directions (say, along rows) the $J_{2}$ bonds are parallel, while
  along the perpendicular axis direction (say, along columns) alternate
  $J_{2}$ bonds are perpendicular to each other, and hence
  form one-dimensional (1D) chevron chains in this direction.  The
  model thus interpolates smoothly between 2D HAFs on the square
  ($\kappa=0$) and triangular ($\kappa=1$) lattices, and also
  extrapolates to disconnected 1D HAF chains ($\kappa \rightarrow
  \infty$).  The classical ($s \rightarrow \infty$) version of the
  model has collinear N\'{e}el order for $0 < \kappa < \kappa_{{\rm
      cl}}$ and a form of noncollinear spiral order for $\kappa_{{\rm
      cl}} < \kappa < \infty$, where $\kappa_{{\rm cl}} =
  \frac{1}{2}$.  For the $s=\frac{1}{2}$ model we use both these
  classical states, as well as other collinear states not realized as
  classical ground-state (GS) phases, as CCM reference states, on top
  of which the multispin-flip configurations resulting from quantum
  fluctuations are incorporated in a systematic truncation hierarchy,
  which we carry out to high orders and then extrapolate to the
  physical limit.  At each order we calculate the GS energy, GS
  magnetic order parameter, and the susceptibilities of the states to
  various forms of valence-bond crystalline (VBC) order, including
  plaquette and two different dimer forms.  We find strong evidence
  that the $s=\frac{1}{2}$ model has two quantum critical points, at
  $\kappa_{c_{1}} \approx 0.72(1)$ and $\kappa_{c_{2}} \approx
  1.5(1)$, such that the system has N\'{e}el order for $0 < \kappa <
  \kappa_{c_{1}}$, a form of spiral order for $\kappa_{c_{1}} <
  \kappa < \kappa_{c_{2}}$ that includes the correct three-sublattice
  $120^{\circ}$ spin ordering for the triangular-lattice HAF at
  $\kappa=1$, and parallel-dimer VBC order for $\kappa_{c_{2}} <
  \kappa < \infty$.
\end{abstract}

%\pacs{Valid PACS appear here}
%\pacs{75.10.Jm}{Quantized spin models}
%\pacs{75.30.Gw}{Magnetic anisotropy}
%\pacs{75.40.-s}{Critical-point effects, specific heats, short-range order (see also 65.40.-b Heat capacities of solids}
%\pacs{75.50.Ee}{Antiferromagnetics}

\maketitle

\section{INTRODUCTION}
\label{intro}
The simultaneous presence of strong frustration and large quantum
fluctuations in highly frustrated and strongly correlated
quantum antiferromagnets on two-dimensional (2D) lattices makes these
systems of huge theoretical interest for investigating possible novel
quantum phases with exotic
ordering.\cite{2D_magnetism_1,2D_magnetism_2,Sachdev:2011} Particular
interest has focussed on the zero-temperature ($T=0$) phase
transitions that can occur between both quasiclassical states showing
various forms of magnetic order and magnetically disordered, quantum
paramagnetic (QP) phases, as some control parameter characterizing
the degree of frustration present in the system, is varied.  The
latter QP phases include both various types of valence-bond
crystalline (VBC) solid and quantum spin-liquid (QSL) states.

Since quantum fluctuations tend to be the largest, other things being
equal, for spins with the smallest spin quantum number $s$, spin-1/2
systems always occupy a special niche.  Furthermore, novel quantum
phases often emerge from the corresponding classical ($s \rightarrow
\infty$) models that exhibit an infinitely degenerate family of ground
states in some region of the classical $T=0$ phase diagram.  What one
typically finds in such a scenario is that this (accidental) classical
ground-state (GS) degeneracy may be lifted entirely (or partially),
by the well-known {\it order by disorder}
mechanism,\cite{Villain:1977} to favor just one (or a few) particular
member(s) of the family as the actual quantum GS phase.  This is often
found to be the case in the quasiclassical limit ($s \gg 1$), where
one works to keep quantum corrections to leading order in the
parameter $1/s$.  However, what is also often then found in those
situations is that for the $s=\frac{1}{2}$ system {\it none} of the
infinitely degenerate set of classical states that form the GS phase
in the $s \rightarrow \infty$ limit survive the quantum fluctuations
to form a stable magnetically ordered GS phase over part, or even all,
of the $T=0$ quantum phase diagram.  Instead, in their place, emerges
one or more novel QP phases with no classical counterpart.

An example of a 2D spin system that fulfills the above scenario is the
anisotropic planar pyrochlore (APP) model (also known as the crossed
chain model).  It comprises a frustrated $J_{1}$--$J_{2}$ Heisenberg
antiferromagnet (HAF) on the 2D checkerboard lattice with
nearest-neighbor (NN) and next-nearest-neighbor (NNN) exchange bonds
of strength $J_{1}>0$ and $J_{2} \equiv \kappa J_{1}>0$, respectively.
It differs from the full $J_{1}$--$J_{2}$ model on the 2D square
lattice by having half of the NNN $J_{2}$ bonds removed, such that
alternate squares have zero or two $J_{2}$ bonds, resulting in a
checkerboard pattern.  It may thus be regarded as a 2D analog of a
three-dimensional anisotropic pyrochlore model of corner-sharing
tetrahedra.

Since the $T=0$ phase diagram of the $s=\frac{1}{2}$ APP model is thus
of considerable interest, much attention has been paid to it, using a
large variety of theoretical
techniques.\cite{Singh:1998,Palmer:2001,Brenig:2002,Canals:2002,
  Starykh:2002,Sindzingre:2002,Fouet:2003,Berg:2003,Tchernyshyov:2003,Moessner:2004,Hermele:2004,
  Brenig:2004,Bernier:2004,Starykh:2005,Schmidt:2006,Arlego:2007,Moukouri:2008,Chan:2011,Bishop:2012_checkerboard}
Despite this huge effort, the structure of its full phase diagram has
remained unsettled, at least until very recently, particularly in the
regime of large values of the frustration parameter, $\kappa > 1$,
which is precisely where the GS phase of the classical ($s \rightarrow
\infty$) version of the model is infinitely degenerate.  For example,
in the limit $\kappa \rightarrow \infty$ the APP model reduces to one
of essentially decoupled one-dimensional (1D) crossed isotropic HAF
chains.  It hence exhibits in that limit a Luttinger QSL GS phase, with
a gapless excitation spectrum of deconfined spin-1/2 spinons.  It
might be supposed (as was argued in
Ref.~\onlinecite{Starykh:2002}) that this
Luttinger-liquid behavior is robust against the gradual turning on of
interchain ($J_{1}$) couplings, so that the 2D system continues to act
as a quasi-1D Luttinger liquid for large but finite values of
$\kappa$.  Such a 2D QSL GS phase is an example of what has been
called a sliding Luttinger liquid (SLL).\cite{Emery:2000,
  Mukhopadhyay:2001,Vishwanath:2001}

Some putative numerical evidence for such a SLL phase in the APP model
at large values of $\kappa$ was claimed in exact diagonalization (ED)
studies\cite{Sindzingre:2002} on small finite-sized lattices of up to
$N=36$ spins.  A more detailed and more careful
analysis\cite{Starykh:2005} of the relevant terms near the 1D
Luttinger liquid fixed point showed, however, that the earlier
prediction\cite{Starykh:2002} of a SLL GS phase was incorrect, and the
same authors\cite{Starykh:2005} suggested that an alternate possible
GS phase in the large-$\kappa$ regime might be the gapped
crossed-dimer valence-bond crystalline (CDVBC) state, with twofold
spontaneous symmetry breaking and without any magnetic order.

A high-order, and numerically very accurate, application of the
coupled cluster method (CCM) (see, e.g.,
Refs.~[\onlinecite{Bishop_1987:ccm,Arponen:1991_ccm,Bi:1991,Bishop:1998,Fa:2004}]
and references cited therein) to the spin-1/2 APP model indeed
showed\cite{Bishop:2012_checkerboard} that the quasiclassical
antiferromagnetic (AFM) state with N\'{e}el ordering is the GS phase
for $\kappa < \kappa_{c_{1}} \approx 0.80 \pm 0.01$, but that the
quantum fluctuations totally destroy the order in {\it all} of the
infinitely degenerate set of AFM states that form the GS phase (for
$\kappa > 1$) in the classical ($s \rightarrow \infty$) case.
Instead, it was found\cite{Bishop:2012_checkerboard} that for $\kappa
> \kappa_{c_{1}}$ there are two stable QP phases in different regimes
of $\kappa$, each with different types of VBC order.  For
$\kappa_{c_{1}} < \kappa < \kappa_{c_{2}} \approx 1.22 \pm 0.02$ the
stable GS phase was found to have plaquette valence-bond crystalline
(PVBC) order, whereas for {\it all} values $\kappa > \kappa_{c_{2}}$
it has crossed-dimer valence-bond crystalline (CDVBC) order.  The
latter CDVBC state has a staggered ordering of dimers along each of
the two sets of crossed $J_{2}$ chains, and hence has twofold
spontaneous symmetry breaking. 

The CCM has also been applied with considerable success to a large
variety of other spin-lattice models (see, e.g.,
Refs.~\onlinecite{Fa:2004,Bishop:1991_ccm,Ze:1998,Kr:2000,Bishop:2000,Fa:2001,schmalfuss,Darradi:2005,Darradi:2008_J1J2mod,Bi:2008_JPCM,
  Bi:2008_PRB_J1xxz_J2xxz,Bi:2009_SqTriangle,richter10,UJack_ccm,Reuther:2011_J1J2J3mod,Farnell:2011,Gotze:2011,Bishop:2012_checkerboard,Li:2012_honey_full,Li:2012_anisotropic_kagomeSq}
and references cited therein) with different types of both
quasiclassical magnetic order and QP order.  These include models that
generalize the $J_{1}$--$J_{2}$ model on the 2D square lattice by
introducing both spatial lattice anisotropy\cite{Bi:2008_JPCM} and
spin anisotropy,\cite{Bi:2008_PRB_J1xxz_J2xxz} as well as several
models that fall in the same half-depleted $J_{1}$--$J_{2}$ class as
the APP model in the sense that they are all obtained from the full
$J_{1}$--$J_{2}$ model on the 2D square lattice by removing half of
the $J_{2}$ bonds in different arrangements.  Examples of this
depleted $J_{1}$--$J_{2}$ square-lattice class are the
($J_{1}$--$J_{2}'$ or) interpolating square-triangle AFM
model\cite{Bi:2009_SqTriangle} and the so-called Union Jack
model.\cite{UJack_ccm}

With respect to an underlying square-lattice geometry the former,
interpolating square-triangle HAF, model contains both NN ($J_{1}$)
bonds and competing NNN ($J_{2}' \equiv \kappa J_{1}$) bonds across
only one of the diagonals of each square plaquette, the same diagonal
in every square.  Each site of the lattice is thus six-connected.
Considered on an equivalent triangular-lattice geometry this model may
be regarded as having two sorts of NN bonds, with $J_{2}'$ bonds along
parallel chains and $J_{1}$ bonds providing an interchain coupling,
such that each triangular plaquette thus contains two $J_{1}$ bonds and
one $J_{2}'$ bond.  By comparison, in the Union Jack model each square
plaquette also has only one NNN ($J_{2}$) bond, but with neighboring
plaquettes now having the $J_{2}$ bond across opposite diagonals.  The
bonds are thus arranged so that on the $2\times 2$ unit cell they form
the pattern of the Union Jack flag, and alternating sites on the
square lattice are thus four-connected and eight-connected.

The corresponding GS phase diagrams of both the spin-1/2 interpolating
square-triangle and the Union Jack HAF models are very different both from each other and from that of the full $J_{1}$--$J_{2}$
square-lattice HAF model.  They {\it both} also differ markedly from
that of the spin-1/2 APP model, despite the fact that both the
interpolating square-triangle and APP models reduce to uncoupled 1D
chains in the $\kappa \rightarrow \infty$ limit.  Presumably the
large-$\kappa$ difference in phase structure has to do with the fact
that whereas the GS phase of the classical ($s \rightarrow \infty$)
version of the APP model is infinitely degenerate after N\'{e}el order
has been destroyed by the effects of frustration, this is not so for
either of the latter models.

\begin{figure*}
\subfigure[]{\scalebox{0.4}{\includegraphics{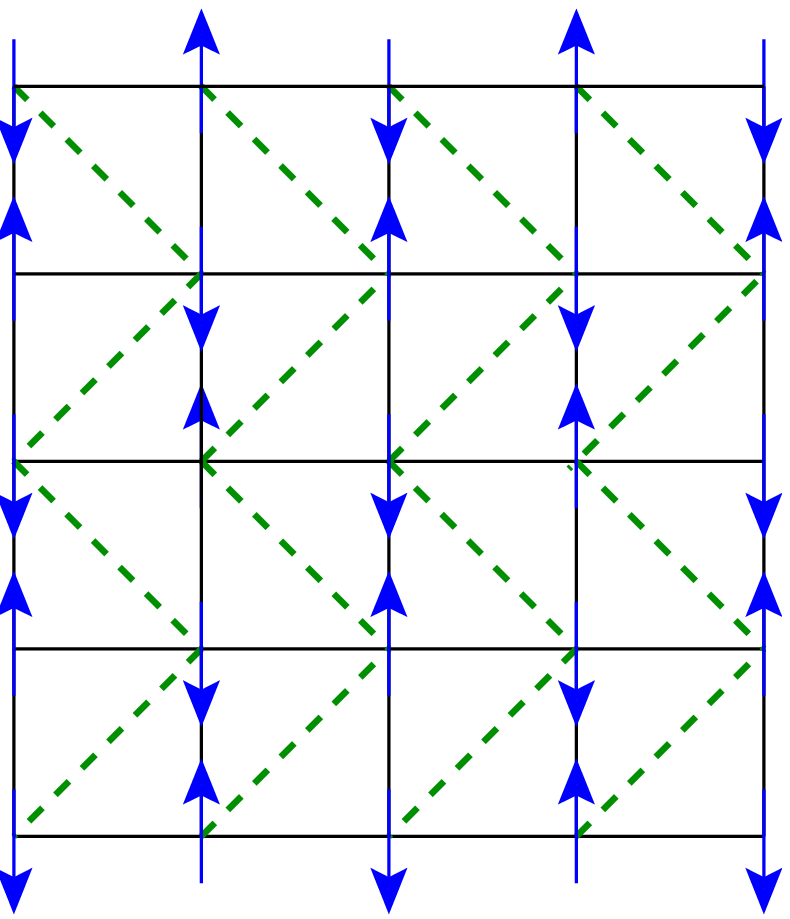}}} \quad
\subfigure[]{\scalebox{0.4}{\includegraphics{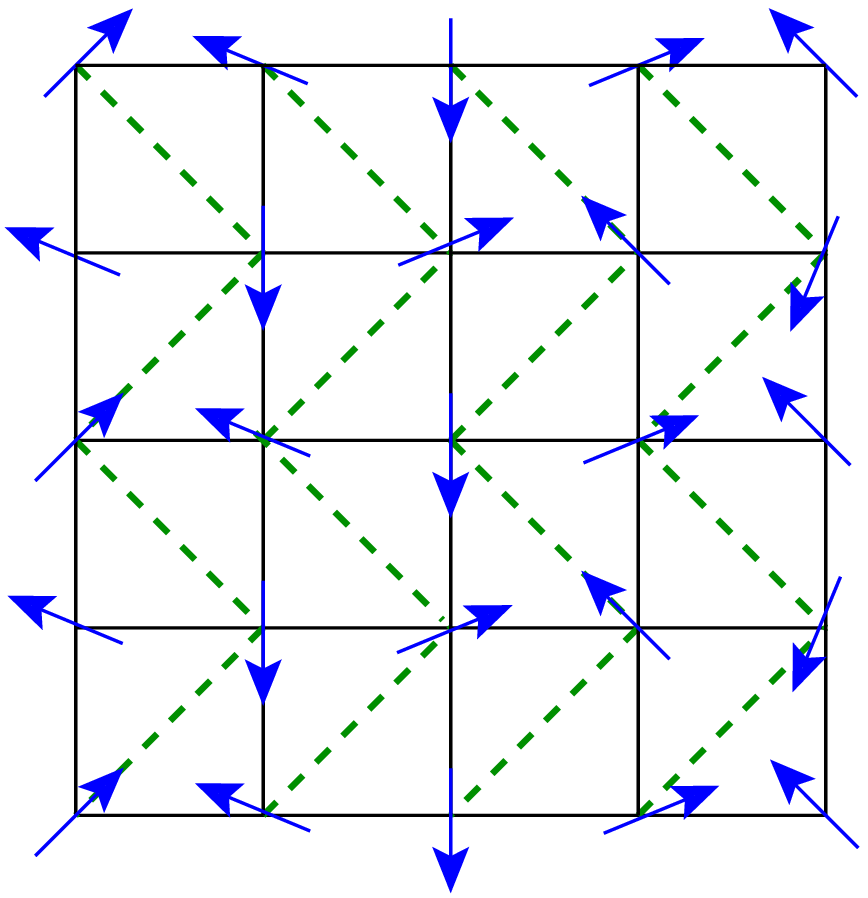}}} \quad
\subfigure[]{\scalebox{0.4}{\includegraphics{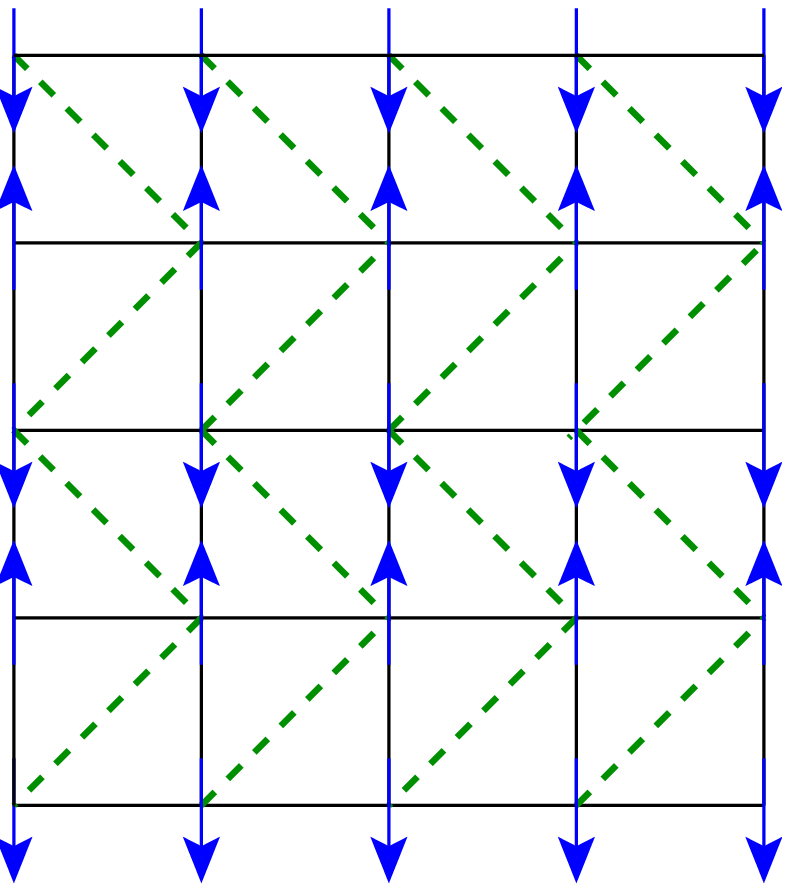}}} \quad
\subfigure[]{\scalebox{0.4}{\includegraphics{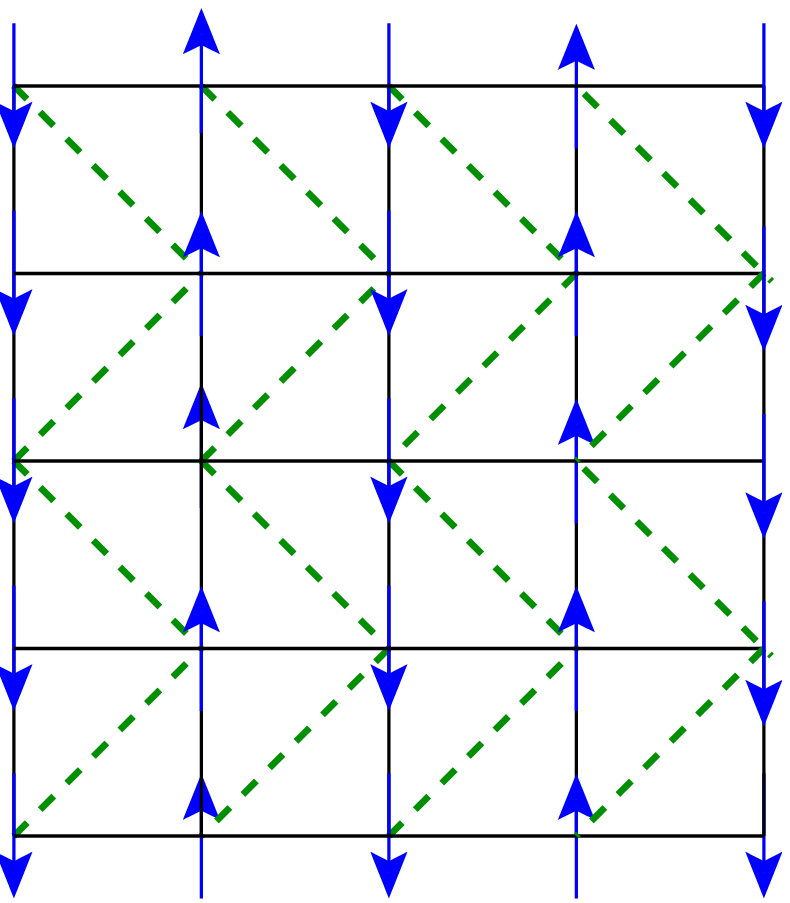}}} \quad
\subfigure[]{\scalebox{0.4}{\includegraphics{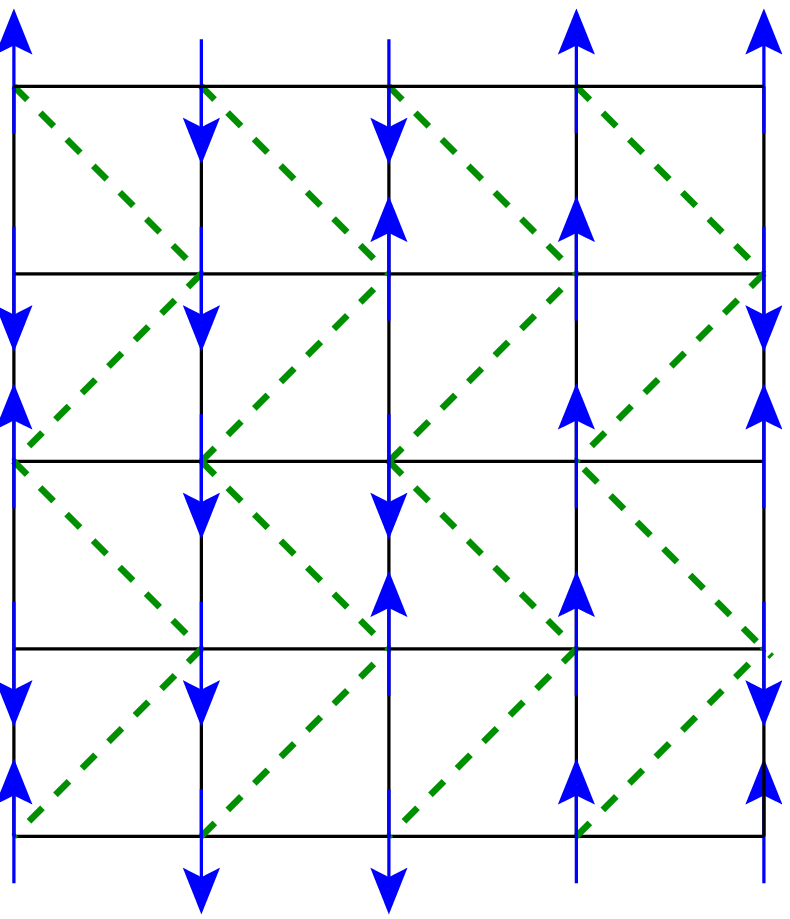}}}
\caption{(Color online) The $J_{1}$--$J_{2}$ chevron-square lattice model
  ($J_{1} \equiv 1$).  The solid (black) lines are $J_{1}$ bonds and
  the dashed (green) lines are $J_{2}$ bonds; (a) N\'{e}el state, (b)
  spiral state (or state s), (c) row-striped state (or state r), (d)
  columnar-striped state (or state c), and (e) doubled alternating
  striped state (or state d).  The blue arrows represent the spins on
  the sites of the chevron lattice.}
\label{model_bonds}
\end{figure*}

There is also another member of the same half-depleted
$J_{1}$--$J_{2}$ class of square-lattice models, which we refer to
here as the chevron-square lattice model, that has largely been ignored up to
now.  Once again, each square plaquette contains only one $J_{2}$
bond, but now in a chevron pattern such that neighboring square
plaquettes along one of the two square-lattice axis directions (say,
rows) have the $J_{2}$ bond on the same diagonal, whereas along the
perpendicular direction (say, columns) neighboring plaquettes have
$J_{2}$ bonds along opposite diagonals.  In our convention
the chevron lattice is henceforth drawn as in Fig.~\ref{model_bonds}
such that the basic chevron stripes of alternating V-shapes and
inverted V-shapes are oriented vertically.
Hence the square plaquettes along each row have the diagonal $J_{2}$
bond oriented in the same direction, and with that direction
alternating between neighboring rows.  On the chevron lattice each
site is thus six-connected, as in the APP and interpolating
square-triangle HAF models.

In view of the fact that other depleted $J_{1}$--$J_{2}$ models on the
square lattice show such a rich variety of GS phase diagrams, the
chevron-square lattice model now seems to be well worth studying
theoretically too for the case $s=\frac{1}{2}$.  The recent
experimental quantum simulation of frustrated magnetism in triangular
optical lattices,\cite{Struck:2011} which exploited the motional
degrees of freedom of ultracold atoms, provides a clear additional
impetus for such studies.  In the experimental
simulation\cite{Struck:2011} a specific modulation of the optical
lattice was used to tune the NN couplings on the triangular lattice
in different directions independently.  By introducing a fast
oscillation of the lattice the experimentalists were thus able to
simulate our previous interpolating square-triangle AFM
model.\cite{Bi:2009_SqTriangle}  Since, as we discuss below in
Sec.~\ref{model_sec}, the $J_{1}$--$J_{2}$ chevron-square lattice model may
also equivalently be regarded as a different form of anisotropic
triangular-lattice HAF, it is conceivable that this model too may be
experimentally realizable with ultracold atoms trapped on a 2D
triangular optical lattice.

In Sec.~\ref{model_sec} we describe the model itself further,
including its classical ($s \rightarrow \infty$) counterpart.  The key
elements of the CCM technique that we will apply to the
$s=\frac{1}{2}$ model are reviewed in Sec.~\ref{ccm_sec}, before
presenting our results in Sec.~\ref{results_sec}.  We end with a brief
summary in Sec.~\ref{summary_sec}.

\section{THE MODEL}
\label{model_sec}
In this paper we study the chevron-square lattice model whose Hamiltonian may
be written as
\begin{equation}
H = J_{1}\sum_{\langle i,j \rangle} \mathbf{s}_{i}\cdot\mathbf{s}_{j} + J_{2}\sum_{\langle\langle i,k \rangle\rangle'}
\mathbf{s}_{i}\cdot\mathbf{s}_{k} \label{H}
\end{equation}
where the operators
$\mathbf{s}_{l}\equiv(s^{x}_{l},s^{y}_{l},s^{z}_{l}$) are the quantum
spin operators on lattice site $l$, with $s^{2}_{l}=s(s+1)$.  We are
interested here in the extreme quantum case $s=\frac{1}{2}$.  On the
square lattice the sum over $\langle i,j \rangle$ runs over all
distinct NN bonds (with exchange coupling strength $J_{1}$), while the
sum over $\langle\langle i,k \rangle\rangle'$ runs over only half of
the distinct NNN diagonal bonds (with exchange coupling strength
$J_{2}$).  In the latter sum only one NNN diagonal bond is retained in
each square plaquette, as arranged in the (vertical) chevron pattern
shown explicitly in Fig.~\ref{model_bonds}.  The primitive unit cell
is thus of size $1 \times 2$.  In both sums in Eq.~(\ref{H}) each bond
is counted once and once only.  In the present paper we consider the
case where both sorts of bonds are AFM in nature, $J_{1}>0$ and $J_{2}
\equiv \kappa J_{1} >0$, and hence act to frustrate one another.
Henceforth, we put $J_{1} \equiv 1$ to set the overall energy scale.

\begin{figure*}
\includegraphics[width=10cm]{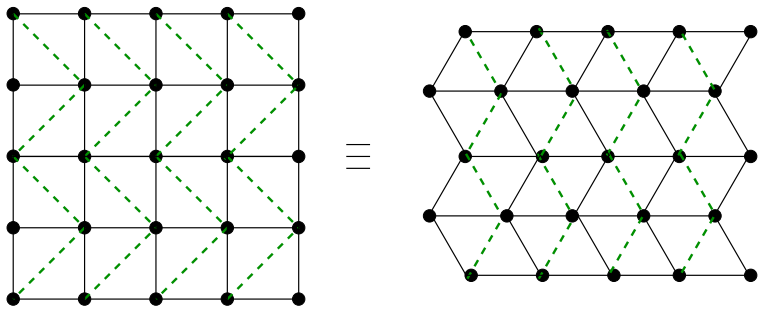}
\caption{(Color online) The $J_{1}$--$J_{2}$ chevron-square lattice model as
  an equivalent anisotropic triangular-lattice HAF model.}
\label{triangle_geom}
\end{figure*}

The model may clearly equivalently be defined on a triangular-lattice
geometry in which every site has six NN sites, four connected to it by
$J_{1}$ bonds and two connected to it by $J_{2}$ bonds, as shown
explicitly in Fig.~\ref{triangle_geom}.
Thus, the chevron-square lattice model may be regarded also as an anisotropic
HAF on a triangular lattice, in which every basic triangular plaquette
contains two NN $J_{1}$ bonds and one NN $J_{2}$ bond in the pattern
shown.  It thus clearly shares some properties with our previous
interpolating square-triangle model\cite{Bi:2009_SqTriangle} in the
sense that both models interpolate smoothly between the HAF on the
square lattice (when the NNN bonds retained on the square lattice are
removed entirely) and the HAF on the triangular lattice (when both NN
and the retained NNN bonds on the square lattice have equal strength).
Both models can be regarded too as anisotropic triangular-lattice HAF
models.  Whereas our former interpolating square-triangle
model\cite{Bi:2009_SqTriangle} had the NN bonds along two of the three
equivalent triangular-lattice directions equal ($J_{1}$) and different
along the third direction ($J_{2}'$), in the current chevron-lattice
model the inequivalent ($J_{2}$) bonds are those along parallel zigzag
(or chevron) chains in one of the three equivalent directions for the
triangular lattice, as shown in Fig.~\ref{triangle_geom}.

The current model clearly also reduces to decoupled 1D chains in the
limit $\kappa \rightarrow \infty$.  Thus, the case $\kappa \gg 1$
corresponds to weakly coupled 1D chains, and the model thus also
interpolates between 1D and 2D limits.

Before considering the extreme quantum limit $s=\frac{1}{2}$ it is
worthwhile to examine first the classical limit, $s \rightarrow
\infty$.  Thus, it is easy to show that the classical chevron-square lattice
model has only two GS phases.  For $\kappa < \frac{1}{2}$ the stable
GS phase has AFM N\'{e}el ordering on the square lattice, as shown in
Fig.~\ref{model_bonds}(a).  At $\kappa = \kappa_{{\rm cl}} =
\frac{1}{2}$ the system undergoes a continuous second-order phase
transition to a state with noncollinear spiral order that persists for
all values $\kappa > \frac{1}{2}$.  The spin direction in this spiral
state, shown in Fig.~\ref{model_bonds}(b), at lattice site ($i,j$)
points at an angle $\alpha_{ij}=\alpha_{0}+[i + \Delta(j)]\alpha$,
where $\Delta(j)$ is defined to be zero if $j$ is even and one if $j$
is odd.  The GS energy per spin in this classical spiral state is thus given by
\begin{equation}
\frac{E}{N} = s^{2}(2\cos\alpha + \kappa\cos2\alpha)\,,
\label{E_classical}
\end{equation}
which is minimized when
\begin{equation}
\alpha = \alpha_{{\rm cl}} = \left\{ 
\begin{array}{rl}
 \pi\,; & \quad \kappa \leq \frac{1}{2} \\
 \cos^{-1}(-\frac{1}{2\kappa})\,; & \quad \kappa > \frac{1}{2}\,.
       \end{array} \right.
\label{Eq_alpha}
\end{equation}

The pitch angle $\phi
\equiv \pi - \alpha$ thus measures the deviation from
N\'{e}el order, and its classical value is
\begin{equation}
\phi_{{\rm cl}} = \left\{ 
\begin{array}{rl}
 0\,; & \quad  \kappa \leq \frac{1}{2} \\
\cos^{-1}(\frac{1}{2\kappa})\,; &  \quad \kappa > \frac{1}{2}\,. 
       \end{array} \right.
\label{Eq_phi_classical}
\end{equation}
It varies from zero for $\kappa < \kappa_{{\rm cl}} =
\frac{1}{2}$ to $\frac{1}{2}\pi$ as $\kappa \rightarrow \infty$ (as
shown later in Fig.~\ref{angle_vs_J2}).  At the isotropic point,
$\kappa = 1$, we regain the classical three-sublattice ordering of the
isotopic triangular-lattice HAF with $\alpha_{{\rm
    cl}}=\frac{2}{3}\pi$.

The GS energy per spin of the classical model is thus given by
\begin{equation}
\frac{E}{Ns^{2}} = \left\{ 
\begin{array}{rl}
 -2+\kappa\,; & \quad  \kappa \leq \frac{1}{2} \\
 -(\frac{1}{2\kappa})-\kappa\,; & \quad \kappa > \frac{1}{2}\,. 
       \end{array} \right.
\label{Eq_E_classical}
\end{equation}
Equation (\ref{Eq_E_classical}) clearly shows that the
classical phase transition at $\kappa = \frac{1}{2}$ is a continuous
(second-order) one, where both the GS energy and its first derivative
with respect to the frustration parameter $\kappa$ are continuous.

In the limit of large $\kappa$ the GS configuration of the classical
chevron-square lattice model represents a set of decoupled 1D HAF chains
(along the chevrons of the model) with sites connected by $J_{2}$
bonds, and with a relative spin orientation approaching $90^{\circ}$
between sites on neighbouring chevrons.  In this classical $\kappa
\rightarrow \infty$ limit there is, however, also complete degeneracy
between all states for which the relative ordering directions of spins
on different HAF chevron chains is arbitrary.  Clearly the spin-1/2
model also similarly reduces in the $\kappa \rightarrow \infty$ limit
to a set of decoupled 1D HAF chevron chains, which exhibit the 1D
Luttinger-liquid behavior described by the exact Bethe ansatz
solution.\cite{Bethe:1931}

In general quantum fluctuations tend to favor states with collinear
spin order over ones with noncollinear order, and it thus seems
possible that at sufficiently large (but finite) values of $\kappa$
the classical spiral order will be destroyed by quantum fluctuations
in the spin-1/2 model of interest here, to yield a state with
collinear spin ordering.  In this case the probable ordering is thus
clearly one with all of the 1D chevron chains exhibiting N\'{e}el
ordering in the same direction, with spins alternating in orientation
(say, up and down).  We note that there is still an infinite family of
such states in which, for example, all of the spins along a given row
(in our convention) can be assigned arbitrarily to be either up or
down.  All of these states are degenerate in energy at the classical
level, with an energy per spin given $E/N = -s^{2}\kappa$.
They include, for example, the three states shown in
Figs.~\ref{model_bonds}(c), \ref{model_bonds}(d), and
\ref{model_bonds}(e) in which, respectively, the spins in a given row
all have the same orientation (e.g.,
$\uparrow\uparrow\uparrow\uparrow\uparrow\cdots$), alternate in
orientation (e.g., $\downarrow\uparrow\downarrow\uparrow\downarrow\cdots$),
and alternate in orientation in a pairwise fashion (e.g.,
$\downarrow\downarrow\uparrow\uparrow\downarrow\cdots$).  Henceforth,
we refer to these three states respectively as the row-striped state (or state r),
the columnar-striped state (or state c), and the doubled alternating striped state (or state d).

Clearly, from Eq. (\ref{Eq_E_classical}), the classical
spiral state (or state s) always has lower energy than this classical
infinitely-degenerate family of collinear AFM states.  Hence, unlike in
the classical APP model, where the corresponding infinitely-degenerate
family of crossed 1D HAF chains {\it does} form the stable GS phase for
{\it all} values $\kappa > 1$, this family never becomes the stable GS
phase for the classical chevron-lattice model except in the $\kappa
\rightarrow \infty$ limit.  Nevertheless, as we have already
suggested, it does form a likely candidate for the stable GS phase of
the $s=\frac{1}{2}$ version of the model considered below, as we
discuss further in Sec.~\ref{results_sec}.  In this case we also
expect that the (accidental) classical degeneracy might again be
lifted by the {\it order by disorder} mechanism.

\section{THE COUPLED CLUSTER METHOD}
\label{ccm_sec}
The
CCM\cite{Bishop_1987:ccm,Arponen:1991_ccm,Bi:1991,Bishop:1998,Fa:2004}
that we use here has become one of the most powerful and most
universally applicable techniques of modern microscopic quantum
many-body theory.  In recent times it has been applied with great
success to a large variety of quantum spin-lattice systems (see, e.g.,
Refs.~\onlinecite{Bishop:1991_ccm,Ze:1998,Kr:2000,Bishop:2000,Fa:2001,schmalfuss,Fa:2004,Darradi:2005,Darradi:2008_J1J2mod,Bi:2008_JPCM,
  Bi:2008_PRB_J1xxz_J2xxz,Bi:2009_SqTriangle,richter10,UJack_ccm,Reuther:2011_J1J2J3mod,Farnell:2011,Gotze:2011,Bishop:2012_checkerboard,Li:2012_honey_full,Li:2012_anisotropic_kagomeSq}
and references cited therein).  In particular, for 2D models of
quantum magnetism, it now provides one of the most accurate methods
available for their study.  Some of its most notable strengths are
that it provides a systematic way to study various candidate GS phases
and their regions of stability, and that in each case the description
is systematically improvable in terms of well-defined hierarchies of
approximations to incorporate more and more of the multispin
configurations present in the exact, fully correlated, GS wave
function.

The CCM formalism is now briefly presented.  We concentrate on its key
ingredients only, as needed for present purposes, and refer the
interested reader to the extensive literature on the method (see,
e.g., Refs.\ \onlinecite{Bishop_1987:ccm,Arponen:1991_ccm,Bi:1991,Bishop:1998,Fa:2004,Ze:1998,Kr:2000,Bishop:2000})
for further details.  Any implementation of the CCM always begins with
the selection of some suitable (normalized) model (or reference) state
$|\Phi\rangle$.  For spin-lattice models this is often conveniently
chosen as a classical state, which may or may not form the actual GS
phase of the classical ($s \rightarrow \infty$) version of the model
in some region of the $T=0$ parameter space.  For the present
chevron-square lattice model we will present below in Sec.\
\ref{results_sec} results based on all five of the states shown in
Fig.\ \ref{model_bonds} as CCM model states, for reasons already
outlined in Sec.\ \ref{model_sec}.  

We denote by $|\Psi\rangle$ and $\langle\tilde{\Psi}|$ respectively the exact
GS ket and bra wave functions of the fully interacting system under
study, where the normalizations are chosen so that
$\langle\tilde{\Psi}|\Psi\rangle=\langle\Phi|\Psi\rangle=\langle\Phi|\Phi\rangle=1$.
The CCM now uses the exponential parametrizations
\begin{equation}
|\Psi\rangle={\rm e}^{S}|\Phi\rangle\,; \quad \langle\tilde{\Psi}|=\langle\Phi|\tilde{S}{\rm e}^{-S}\,,   \label{expon_parametration}
\end{equation}
to incorporate explicitly the multispin-flip configurations in
$|\Psi\rangle$ and $\langle\tilde{\Psi}|$ above and beyond those
contained in the chosen model state $|\Phi\rangle$.  The ket- and bra-state
correlation operators are then expressed as
\begin{equation}
S=\sum_{I\neq0}{\cal S}_{I}C^{+}_{I}\,; \quad \tilde{S}=1+\sum_{I\neq0}{\cal \tilde{{S}}}_{I}C^{-}_{I}\,,    \label{correlation_operator}
\end{equation}
where, by definition, $C^{+}_{0} \equiv 1$ is the identity operator,
and $C^{-}_{I}|\Phi\rangle=0=\langle\Phi|C^{+}_{I}, \forall I \neq 0$.
The set-index $\{I\}$ represents a multispin-flip configuration with respect to
state $|\Phi\rangle$, as we elaborate further below.  The operator
$C^{+}_{I}$ ($\equiv (C^{-}_{I})^{\dagger}$) thus represents a
multispin-flip creation operator with respect to $|\Phi\rangle$ considered
as a generalized vacuum state.

It is now convenient to choose a set of local coordinate frames in
spin space defined for each model state separately, such that on each
lattice site the spin aligns along the negative $z$ axis (downwards).  Such
rotations obviously leave the basic SU(2) spin commutation relations
unchanged.  In such a basis the $C^{+}_{I}$ operators now have the
universal form, $C^{+}_{I} \equiv s^{+}_{j_{1}}s^{+}_{j_{2}}\cdots
s^{+}_{j_{n}}$ in terms of the usual single-spin raising operators
$s^{+}_{j} \equiv s^{x}_{j}+is^{y}_{j}$.  For a spin $s$ the raising
operator $s^{+}_{j}$ may be applied a maximum of 2$s$ times on a given
site $j$, and hence the set-index $\{I\} \equiv
\{j_{1},j_{2},\cdots,j_{n};\, n=1,2,\cdots\}$ where any given site-index
$j_{k}$ may appear a maximum of 2$s$ times.  In the present
$s=\frac{1}{2}$ case therefore, no index may be repeated.

The ket- and bra-state coefficients $\{{\cal S}_{I},\tilde{\cal S}_{I}\}$, which
from Eqs.\ (\ref{expon_parametration}) and
(\ref{correlation_operator}) completely specify the GS wave function,
may now be obtained by minimizing the GS energy functional,
$\bar{H}=\bar{H}({\cal S}_{I},\tilde{\cal
  S}_{I})\equiv\langle\tilde{\Psi}|H|\Psi\rangle$, where $H$ is the
Hamiltonian of the system, with respect to all of the coefficients
$\{\tilde{\cal S}_{I}\}$ and $\{{\cal S}_{I}\}$ separately, $\forall I \neq 0$.  Use
of Eqs.\ (\ref{expon_parametration}) and (\ref{correlation_operator})
then leads simply to the respective coupled sets of equations
$\langle\Phi|C^{-}_{I}{\rm e}^{-S}H{\rm e}^{S}|\Phi\rangle = 0$ and
$\langle\Phi|\tilde{{\cal S}}{\rm e}^{-S}[H,C^{+}_{I}]{\rm
  e}^{S}|\Phi\rangle = 0; \forall I \neq 0$, which are fully
equivalent to the GS ket and bra Schr\"{o}dinger equations
$H|\Psi\rangle = E|\Psi\rangle$ and $\langle\tilde{\Psi}|H =
E\langle\tilde{\Psi}|$.  The latter set of equations may equivalently
be written as $\langle\Phi|\tilde{S}({\rm e}^{-S}H{\rm
  e}^{S}-E)C^{+}_{I}|\Phi\rangle = 0, \forall I \neq 0$.

Once these sets of CCM equations have been solved for the sets of
correlation coefficients $\{{\cal S}_{I}\}$ and $\{\tilde{\cal
  S}_{I}\}$, all GS quantities may now be calculated in terms of them.
The GS energy is unique in that its evaluation requires only the
ket-state coefficients, $E=\langle\Phi|{\rm e}^{-S}H{\rm
  e}^{S}|\Phi\rangle$, whereas any other GS quantity requires the
bra-state coefficients also.  For present purposes, for example, we
will also calculate the magnetic order parameter, defined to be the
local average on-site magnetization, $M$, in the rotated spin
coordinates,
$M \equiv -\frac{1}{N}\langle\tilde{\Psi}|\sum^{N}_{j=1}s^{z}_{j}|\Psi\rangle$,
where $N$ is the number of lattice sites.  A key feature of the CCM
exponential parametrizations of Eq.\ (\ref{expon_parametration}) is
that the method automatically satisfies the Goldstone linked cluster
theorem, even when truncations are made in the multispin
configurations $\{I\}$ retained in Eq.\ (\ref{correlation_operator}).
Thus, the method is automatically size-extensive at every level of
approximation, and we may take the infinite-lattice limit, $N
\rightarrow \infty$, from the very outset.  Similarly, the method also
satisfies the Hellmann-Feynman theorem at all levels of truncation.

It is important to note too that while the CCM equations are
intrinsically nonlinear, due to the presence of the basic correlation
operator $S$ in the exponentiated form ${\rm e}^{S}$, it only ever
appears in the equations to be solved in the similarity transform,
${\rm e}^{-S}H{\rm e}^{S}$, of the Hamiltonian.  By making use of the
well-known nested commutator expansion for ${\rm e}^{-S}H{\rm e}^{S}$,
it is readily seen that the basic SU(2) spin commutation relations
imply that the otherwise infinite series of nested commutators
actually terminates {\it exactly} at second-order terms in $S$ for
Hamiltonians of the form of Eq.\ (\ref{H}) used here (and see, e.g.,
Refs.\ \onlinecite{Fa:2004,Ze:1998} for further details).  A similar
exact termination generally also applies to the evaluation of the GS
expectation value of any operator, such as $M$ above, that we calculate.

Thus, the CCM formalism is exact if all multispin-flip configurations are
included in the index-set $\{I\}$, and the equations that we need to
solve in practice are coupled sets of multinomial equations for the
coefficients $\{{\cal S}_{I}\}$ and linear equations for the
coefficients $\{\tilde{\cal S}_{I}\}$, in which the solutions for
$\{{\cal S}_{I}\}$ are needed as input.  Of course, in practice we
need to make finite-size truncations in the multispin-flip configurations
retained in the indices $\{I\}$.  We employ here the localized
(lattice-animal-based subsystem) LSUB$n$ scheme,\cite{Fa:2004,Ze:1998}
which has been well-studied and greatly utilized for a wide variety of
spin-1/2 lattice
systems.\cite{Bishop:1991_ccm,Ze:1998,Kr:2000,Bishop:2000,Fa:2001,schmalfuss,Fa:2004,Darradi:2005,Darradi:2008_J1J2mod,Bi:2008_JPCM,
  Bi:2008_PRB_J1xxz_J2xxz,Bi:2009_SqTriangle,richter10,UJack_ccm,Reuther:2011_J1J2J3mod,Farnell:2011,Gotze:2011,Bishop:2012_checkerboard,Li:2012_honey_full,Li:2012_anisotropic_kagomeSq}
At the {\it n}th level of approximation in the LSUB$n$ scheme all
possible multispin-flip configurations are retained in the index-set
$\{I\}$ over different locales on the lattice defined by $n$ or fewer
contiguous sites.  In other words, all lattice animals of size up to
$n$ sites are populated with flipped spins (with respect to the model
state $|\Phi\rangle$ as reference state) in all possible ways.  Such
animals (or contiguous clusters) are defined to be contiguous in this
sense if every site in the cluster is adjacent (in the NN sense) to at
least one other site in the cluster.  For our present model we define
NN pairs on the triangular-lattice geometry (as shown in Fig.\
\ref{triangle_geom}), rather than on the square-lattice geometry,
since we wish to treat all pairs connected by both $J_{1}$ and $J_{2}$
bonds on an equal footing.

The number $N_{f}$ of such fundamental configurations, which are
distinct under the space- and point-group symmetries of the lattice
and of the model state being used, increases very rapidly with the
truncation index $n$ of the LSUB$n$ hierarchy.  In the present study,
by making use of a highly efficient parallelized CCM code\cite{cccm}
and supercomputer resources we have been able to perform LSUB$n$
calculations up to the LSUB10 level for each of the four collinear
model states of Figs.\ \ref{model_bonds}(a), (c), (d), and (e), and up
to the LSUB8 level for the spiral state of Fig.\ \ref{model_bonds}(b).
For example, using the triangular-lattice geometry, $N_{f}=541578$
for the LSUB10 approximation using the row-striped state of Fig.\
\ref{model_bonds}(c) as CCM model state.

Although we never need to perform any finite-size scaling of our
results, since we work from the outset in the thermodynamic limit, $N
\rightarrow \infty$, we do need to extrapolate our approximate LSUB$n$
sequence for any physical quantity to the exact $n \rightarrow \infty$
limit.  For example, for the GS energy per spin, $E/N$, we use the
well-tested extrapolation scheme~\cite{Kr:2000,Bishop:2000,Fa:2001,Darradi:2005,schmalfuss,Bi:2008_PRB_J1xxz_J2xxz,Darradi:2008_J1J2mod,
  Bi:2008_JPCM,richter10,Reuther:2011_J1J2J3mod,Bishop:2012_checkerboard,Li:2012_anisotropic_kagomeSq}
%%%%%%%%%%%%
\begin{equation}
E(n)/N = a_{0}+a_{1}n^{-2}+a_{2}n^{-4}\,.     \label{E_extrapo}
\end{equation}
%%%%%%%%%%%%%
As is to be expected other physical quantities do not converge as
rapidly as the energy.  For example, for the magnetic order parameter,
$M$, we generally find that a scaling law with leading power $1/n$, i.e.,
%%%%%%%%%%%%%%%
\begin{equation}
M(n) = b_{0}+b_{1}n^{-1}+b_{2}n^{-2}\,,    \label{M_extrapo_standard}
\end{equation}
%%%%%%%%%%%%%%
works well for most systems with even moderate amounts of
frustration.~\cite{Kr:2000,Bishop:2000,Fa:2001,Darradi:2005} However,
for systems very close to a QCP or for which the magnetic order
parameter of the phase under study is very small or zero, the scaling
law of Eq.\ (\ref{M_extrapo_standard}) has been found to overestimate
the magnetic order and to predict a too large value for the critical
value of the the frustration parameter that is driving the transition.
In such cases we use the well-studied extrapolation
scheme\cite{Darradi:2005,schmalfuss,Bi:2008_JPCM,Bi:2008_PRB_J1xxz_J2xxz,Darradi:2008_J1J2mod,richter10,Reuther:2011_J1J2J3mod,Bishop:2012_checkerboard,Li:2012_anisotropic_kagomeSq}
%%%%%%%%%%%%%
\begin{equation}
M(n) = c_{0}+c_{1}n^{-1/2}+c_{2}n^{-3/2}\,.   \label{M_extrapo_frustrated}
\end{equation}
%%%%%%%%%%%%%%%

Naturally, for any physical quantity $Q$ we may always test for the correct leading power in the corresponding LSUB$n$ extrapolation scheme by first fitting to a form
\begin{equation}
Q(n) = q_{0}+q_{1}n^{-\nu}\,,    
\label{M_extrapo_nu}
\end{equation}
where the exponent $\nu$ is also a fitting parameter.  For the GS
energy we generally find a value $\nu \approx 2$ for a wide variety of
both unfrustrated and (strongly) frustrated systems, as is also the
case here, as we discuss in Sec.\ \ref{results_sec} in more detail.  On
the other hand, for the magnetic order parameter, $M$ we find $\nu
\approx 1$ for many systems, even moderately frustrated systems such
as the triangular-lattice HAF, which is the special case $\kappa=1$ of
the present model.  However, in situations where we are either close
to a transition or where $M$ is (close to) zero for the phase under
study we generally find $\nu \approx 0.5$.  Both cases are present in
the current model, as we discuss more fully in Sec.\
\ref{results_sec}.  Once such values for the leading exponent have
been found, they provide the subsequent rationale for the use of Eqs.\
(\ref{E_extrapo})--(\ref{M_extrapo_frustrated}), for example.

\section{RESULTS}
\label{results_sec}
We report here on our CCM calculations for the present spin-1/2
$J_{1}$--$J_{2}$ chevron-square lattice model of Eq.~(\ref{H}) based in turn
on each of the N\'{e}el, spiral, row-striped, columnar-striped, and
doubled alternating striped states shown, respectively in
Figs.~\ref{model_bonds}(a,b,c,d,e).  The computational power at our
disposal is such that we can perform LSUB$n$ calculations for each of
the collinear model states with $n \leq 10$, but for the more exacting
spiral model state only for $n \leq 8$.

In the first place we discuss our results obtained using the spiral
model state.  Whereas the classical ($s \rightarrow \infty$) version
of the model has a second-order phase transition from N\'{e}el order
(for $\kappa < \kappa_{{\rm cl}}$) to spiral order (for $\kappa >
\kappa_{{\rm cl}}$) at the value $\kappa = \kappa_{{\rm cl}} =0.5$ of
the frustration parameter, our CCM results show a shift of this
critical point to a value $\kappa=\kappa_{{\rm cl}}\approx 0.72$ for
the present $s=\frac{1}{2}$ quantum case.  This finding is in complete
agreement with the general observation, which is found to be true for
a wide variety of different spin-lattice models, that quantum
fluctuations generally favor collinear over noncollinear forms of spin
ordering, as discussed previously in Sec.\ \ref{model_sec}.

\begin{figure}[t]
\hspace{-0.3cm}\includegraphics[angle=270,width=9cm]{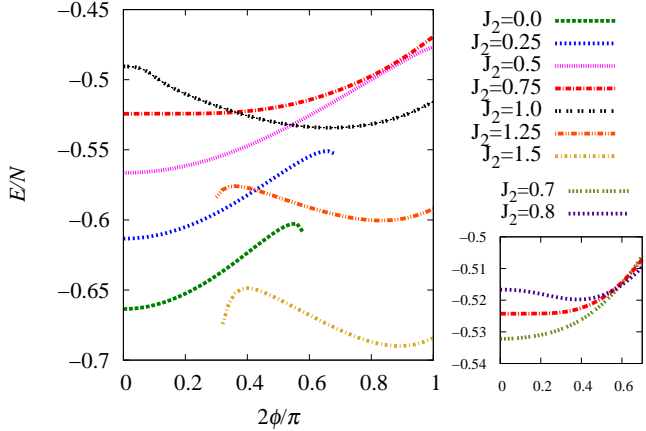}
\caption{(Color online) Ground-state energy per spin, $E/N$, as a
  function of the spiral angle $\phi$, of the spin-1/2
  $J_{1}$--$J_{2}$ chevron-square lattice model ($J_{1} \equiv 1$,
  $J_{2}>0$).  We show results using the LSUB4 approximation of the
  CCM based on the spiral model state for some illustrative values of
  $J_{2}$ in the range $0 \leq J_{2} \leq 1.5$.  For $J_{2} \lesssim
  0.752$ the minimum is at $\phi=0$ (N\'{e}el order), whereas for
  $J_{2} \gtrsim 0.752$ the minimum occurs at $\phi=\phi_{{\rm
      LSUB}4}\neq0$, signalling a phase transition at $J_{2} \approx
  0.752$ in this approximation.}
\label{E_vs_angle}
\end{figure}

\begin{figure}[!t]
  \includegraphics[width=6cm,angle=270]{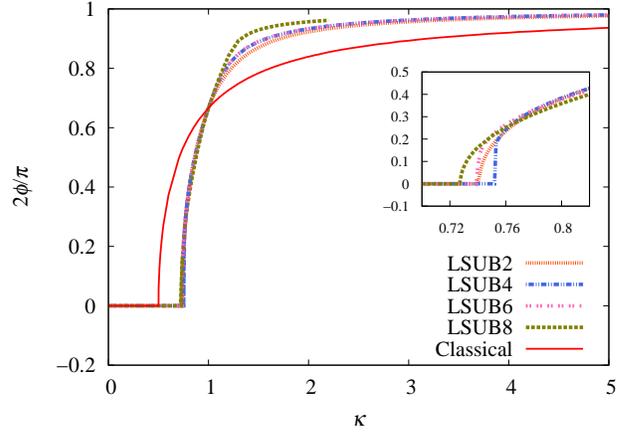}
  \caption{The angle $\phi_{{\rm LSUB}n}$ that minimizes the energy
    per spin, $E_{{\rm LSUB}n}(\phi)/N$, of the spin-1/2
    $J_{1}$--$J_{2}$ chevron-square lattice model (with $J_{1} > 0$) versus
    the frustration parameter $\kappa \equiv J_{2}/J_{1}$, in the
    LSUB$n$ approximations with $n=\{2,4,6,8\}$ based on the spiral
    model state.  The corresponding classical result, $\phi_{{\rm
        cl}}$, of Eq.\ (\ref{Eq_phi_classical}) is also shown for
    comparison.}
\label{angle_vs_J2}
\end{figure}

Curves such as those shown in Fig.\ \ref{E_vs_angle} show clearly that
the CCM results based on the N\'{e}el model state (with $\phi=0$) give
the minimum GS energy for all values of $\kappa < \kappa_{c_{1}}$,
where our estimate for $\kappa_{c_{1}}$ is itself dependent on the
level of LSUB$n$ approximation used, as may be seen from Fig.\
\ref{angle_vs_J2}.  
%%%%%%%%%%%%
The general shape of the curves shown in Fig.\
\ref{E_vs_angle} is broadly similar to that of their classical
counterparts, $E/N = s^{2}(\kappa\cos 2\phi - 2\cos \phi$), from Eq.\
(\ref{E_classical}), but with several important differences.  One is
that the crossover from the minimum being at $\phi = 0$ (N\'{e}el
order) to a value $\phi > 0$ (spiral order) occurs at an appreciably
higher value $\kappa_{c_{1}} \approx 0.72$ in the $s=\frac{1}{2}$ case
than at the value $\kappa_{{\rm cl}}=0.5$ for the classical ($s
\rightarrow \infty$) case.  Another difference is that the crossover
curve itself becomes even flatter in the spin-1/2 case than in the
classical case.

Thus, a close inspection of curves such as those displayed in Fig.\
\ref{E_vs_angle} for the LSUB4 case shows that, at this level of
approximation, for example, for all values $\kappa \lesssim 0.752$ the
only minimum in the GS energy is at $\phi = 0$.  When this value is
approached asymptotically from below the LSUB4 energy curves become
very flat around $\phi=0$, indicating the disappearance at $\phi=0$
not only of the second derivative ${\rm d}^{2}E/{\rm d}\phi^{2}$ but
probably also of one or more of the higher derivatives ${\rm d}^{n}E/{\rm
  d}\phi^{n}$ with $n>2$ (as well as the first derivative ${\rm
  d}E/{\rm d}\phi$).  Similar curves occur for other LSUB$n$
approximations.  By contrast, in the classical case, at $\kappa =
\kappa_{{\rm cl}}$, ${\rm d}^{n}E/{\rm d}\phi^{n}$ vanishes only for
$n \leq 2$.  For values $\kappa > \kappa_{{\rm cl}}$ the classical
curves now have a maximum at $\phi = 0$ and a minimum at the value
$\phi_{{\rm cl}}$ given by Eq.\ (\ref{Eq_phi_classical}).  This
behavior is broadly echoed in the LSUB4 $s=\frac{1}{2}$ curves shown
in Fig.\ \ref{E_vs_angle} for $\kappa > \kappa_{c_{1}}$.  

However, we note from Fig.\ \ref{angle_vs_J2} (and see especially the
inset) that the crossover from one minimum ($\phi = 0$, N\'{e}el)
solution to the other ($\phi \neq 0$, spiral) one appears to be rather
abrupt, particularly for the LSUB$n$ approximations with $n=4,6$, with
the pitch angle rising very sharply from a zero value on the N\'{e}el
side to a nonzero value on the spiral side.  This very sharp rise is
itself just a reflection of the extreme flatness of the corresponding
GS energy curves in Fig.\ \ref{E_vs_angle} near the crossover point
$\kappa = \kappa_{{\rm cl}}$.  This very flatness makes it difficult
for us to distinguish between the two scenarios of a discontinuous
jump in pitch angle versus a continuous but very steep rise in pitch
angle as we traverse the transition in the frustration parameter
$\kappa$ from the N\'{e}el phase into the spiral one.  These two
scenarios would correspond, respectively, to a weakly first-order
transition versus a (continuous) second-order one.  More compelling
evidence for both the numerical value $\kappa_{c_{1}}$ and the nature
of the N\'{e}el-spiral quantum phase transition comes from our
results, presented below, for the local on-site magnetization (which
provides the relevant order parameter for the transition).

However, before doing so, we comment further on Figs.\
\ref{E_vs_angle} and \ref{angle_vs_J2}.  Firstly, we see from Fig.\
\ref{E_vs_angle} that all of the CCM LSUB$n$ approximation give
exactly the classical value $\phi = \frac{1}{3}\pi$ for the pitch
angle at the value $\kappa =1$ corresponding to the spin-1/2 HAF on
the triangular lattice.  This corresponds to the correct $120^{\circ}$
three-sublattice ordering demanded by symmetry in this case.  The fact
that every LSUB$n$ approximation preserves this symmetry is a simple
consequence of us defining the fundamental LSUB$n$ configurations on
the triangular-lattice geometry shown in Fig.\ \ref{triangle_geom} (in
which sites connected by both $J_{1}$ and $J_{2}$ bonds are defined to
be NN pairs).  By contrast, had we defined the fundamental LSUB$n$
configurations on the underlying square-lattice geometry (in which
only sites connected by $J_{1}$ bonds are NN pairs, and where sites
connected by $J_{2}$ bonds are NNN pairs), the exact symmetry at
$\kappa = 1$ would only be approximately preserved in any LSUB$n$
approximation with a finite value of $n$.

We also note from Fig.\ \ref{angle_vs_J2} that when $\kappa > 1$ the
pitch angle $\phi$ approaches the asymptotic ($\kappa \rightarrow
\infty$) limiting value $\frac{1}{2}\pi$ much faster for the quantum
$s=\frac{1}{2}$ model than for its classical ($s \rightarrow \infty$)
counterpart.  This is a first direct piece of evidence that quantum
fluctuations favor, in this limit, collinear ordering along the weakly
coupled 1D chevron chains.  Figure \ref{angle_vs_J2} also shows that
the CCM LSUB8 approximation based on the spiral model state appears
slightly anomalous in the region $\kappa > 1$, with the solution even
becoming unstable for values $\kappa \gtrsim 2.2$.  We take this as a
first indication that the quantum spiral state itself loses its
stability as the actual GS phase in this regime, as we discuss much
more fully below.

Before doing so, however, we note first from Fig.\ \ref{E_vs_angle}
that our CCM solutions at a given LSUB$n$ level of implementation
(viz., LSUB4 in Fig.\ \ref{E_vs_angle}) exist only for certain ranges
of the spiral pitch angle $\phi$ for some values of $\kappa$.  Thus,
for example, for the case $\kappa = 0$ pertaining to the pure
square-lattice HAF (with NN interactions only), the CCM LSUB4 solution
based on the spiral model state of Fig.\ \ref{model_bonds}(b) only
exists for values of the frustration parameter in the range $0 \leq
\kappa \lesssim 0.29\pi$.  For this value $\kappa=0$, where the
N\'{e}el state is the physical GS phase, it seems that any attempt to
drive the system too far away from collinearity leads to the CCM
solutions themselves becoming unstable in the sense that a real
solution simply ceases to exist.  Similarly, for example, when
$\kappa=1.5$, Fig.\ \ref{E_vs_angle} shows that the CCM LSUB4 solution
exists only for values $0.15\pi \lesssim \kappa \lesssim 0.5\pi$.  In
this case the spiral state provides the stable GS phase, and any
attempt now to move too close to collinearity leads again to
instability.  

Similar instabilities, which manifest themselves as
termination points, occur in all LSUB$n$ approximation (with $n>2$).
Indeed, such terminations of CCM solutions are both very common and
well understood (see, e.g., Refs.\
\onlinecite{Fa:2004,Bi:2009_SqTriangle}).  They are always reflections
of the actual quantum phase transitions in the system under study.  In
the present case this is simply the transition between the N\'{e}el
and spiral phases.

\begin{figure}
  \includegraphics[angle=270,width=8cm]{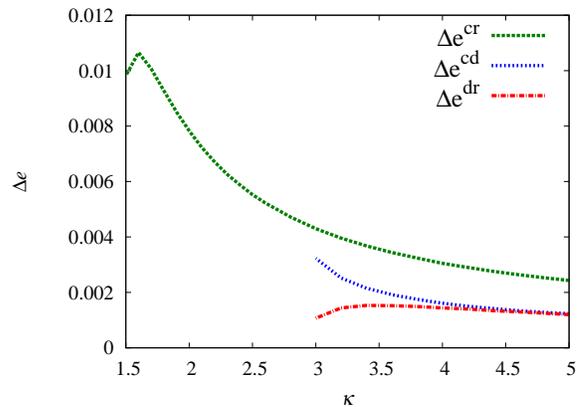}
  \caption{(Color online) Various GS energy differences, $\Delta
    e^{\rm ab} \equiv e^{{\rm state\; a}} - e^{{\rm state\; b}}$, with
    $e \equiv E/N$, between states a and b, versus the frustration
    parameter $\kappa \equiv J_{2}/J_{1}$, for the spin-1/2
    chevron-square lattice model (with $J_{1}>0$), where states a and b
    are chosen to be the states r (row-striped), c (columnar-striped), and
    d (doubled alternating striped) of Figs.\
    \ref{model_bonds}(c)--(e), respectively.  In each case the energy
    per spin, $e$, is the LSUB$\infty$ result obtained from Eq.\
    (\ref{E_extrapo}) using CCM LSUB$n$ results with
    $n=\{4,6,8,10\}$.}
\label{E_diff}
\end{figure}

  In view of the fact that we have indications or suggestions that the
  classical spiral phase might become unstable (for larger values of
  the frustration parameter $\kappa$) against quantum fluctuations, we
  have also performed CCM calculations based on the three other
  collinear states shown in Fig.\ \ref{model_bonds}(c),
  \ref{model_bonds}(d), and \ref{model_bonds}(e).  All of these share
  the feature that the magnetic ordering along each chevron chain
  (i.e., sites joined by $J_{2}$ bonds) is that of a 1D N\'{e}el HAF.
  As explained in Sec.\ \ref{model_sec}, there is actually an infinite
  family of such states, all of which are degenerate in energy at the
  classical level.  As expected, Fig.\ \ref{E_diff} now shows that
  this accidental classical degeneracy is lifted by quantum
  fluctuations.
\begin{figure*}[!t]
\mbox{
\subfigure[]{\scalebox{0.3}{\includegraphics[angle=270]{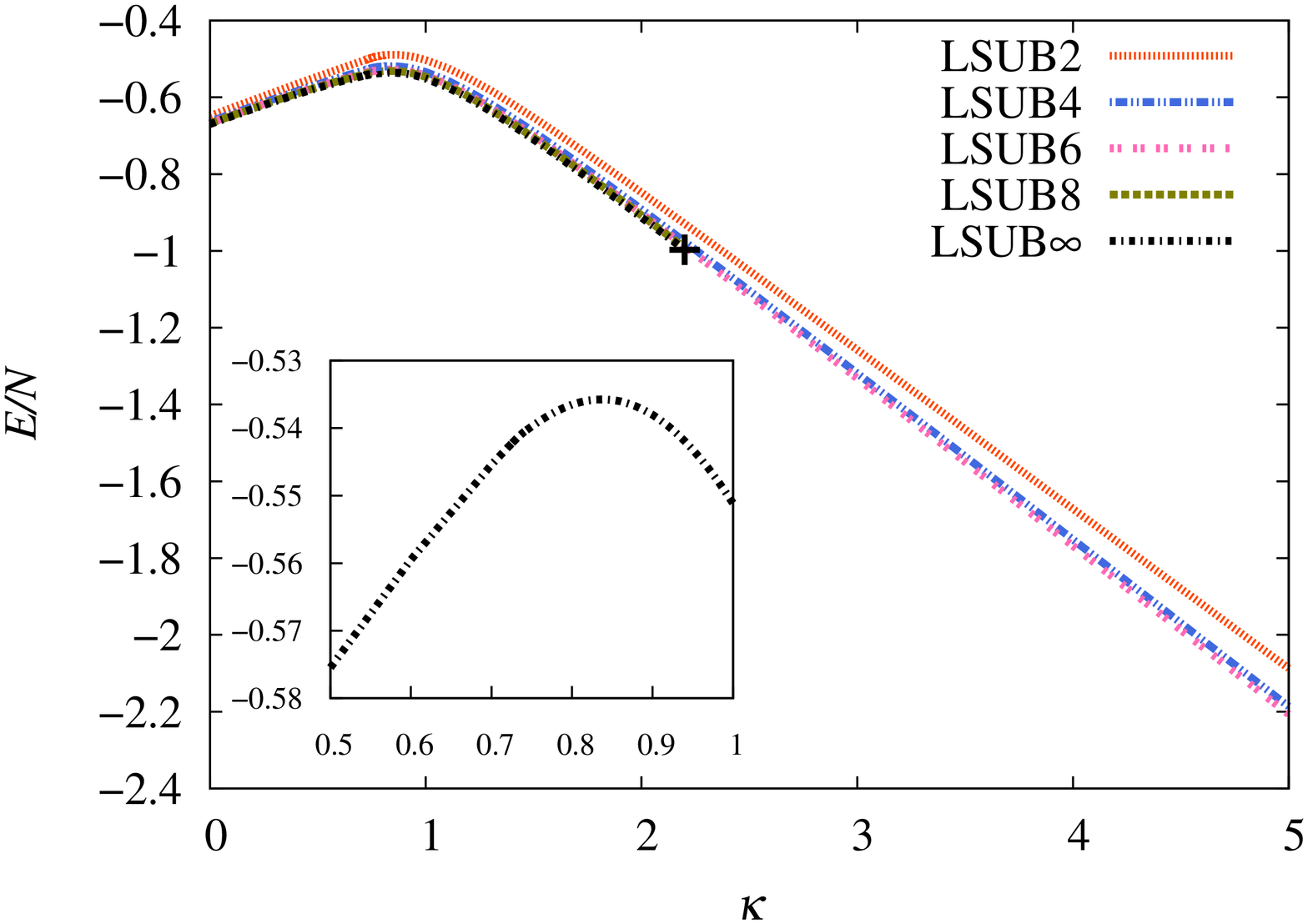}}}
\subfigure[]{\scalebox{0.3}{\includegraphics[angle=270]{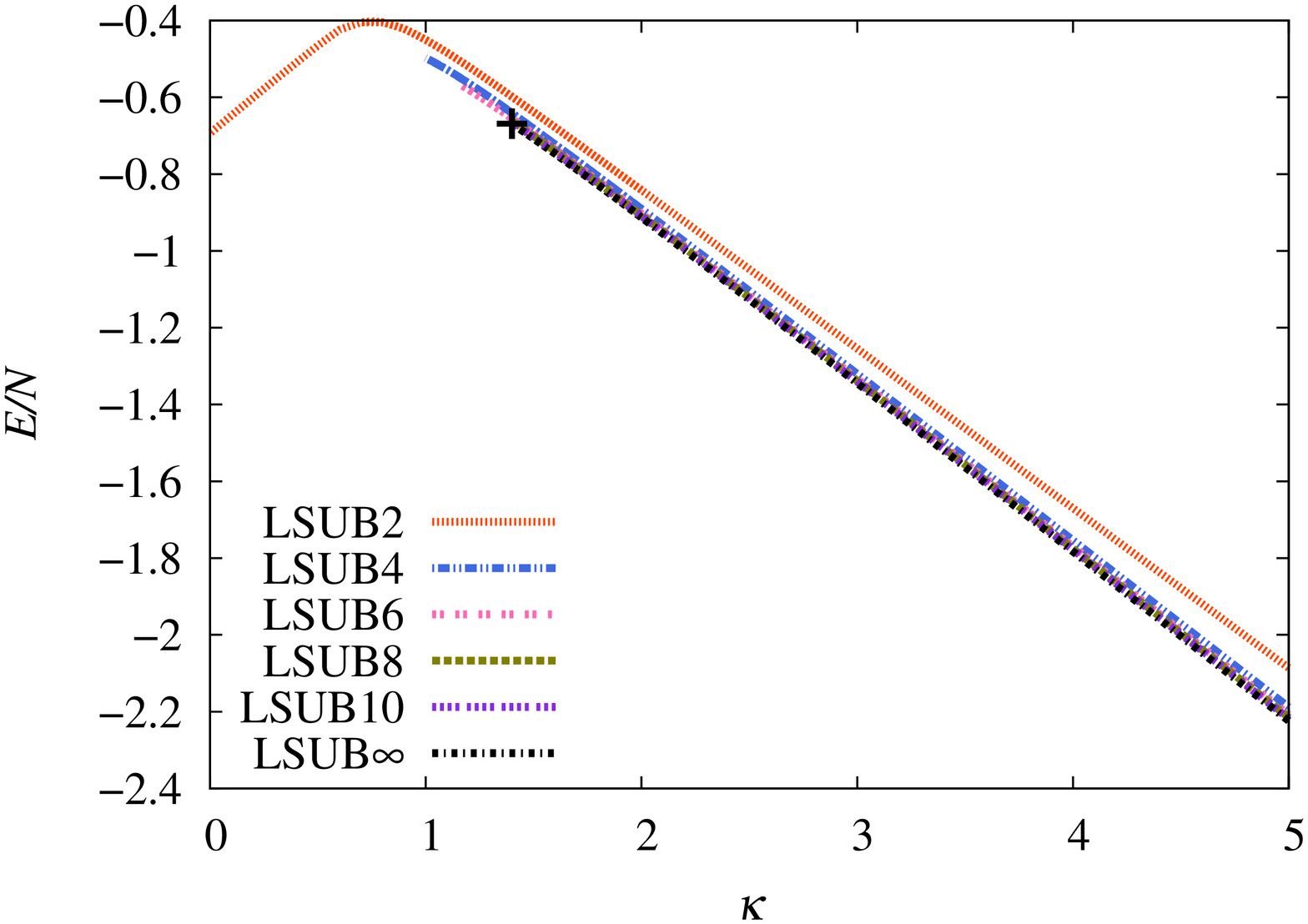}}}
}
\caption{(Color online) CCM LSUB$n$ results for the GS energy per
  spin, $E/N$, versus the frustration parameter $\kappa \equiv
  J_2/J_{1}$, for the spin-1/2 $J_{1}$--$J_{2}$ chevron-square lattice model
  (with $J_{1}>0$), using (a) the N\'{e}el and spiral states of Figs.\
  \ref{model_bonds}(a) and \ref{model_bonds}(b) respectively as model
  states with $n=\{2,4,6,8\}$, and (b) the row-striped state of Fig.\
  \ref{model_bonds}(c) as model state, with $n=\{2,4,6,8,10\}$.  In
  each case we also show the extrapolated LSUB$\infty$ result
  obtained from Eq.~(\ref{E_extrapo}) using the data set
  $n=\{2,4,6,8\}$. }
\label{E}
\end{figure*}
As a parenthetic note we add that in Fig.\ \ref{E_diff} the results
using state d as CCM model state are shown only in the range $\kappa
\geq 3$, since real solutions for all of the corresponding LSUB$n$
approximations used (viz., $n=\{4,6,8,10\}$) exist only in this
range for this state.  Corresponding terminations for states c and r
occur only for values $\kappa < 1.5$, as we discuss more fully below
for state r.  Although the differences in the energies between the
three states are small, it is clear that the row-striped state of
Fig.\ \ref{model_bonds}(c) has the lowest GS energy for all values of
the frustration parameter, $\kappa$, shown.  Henceforth, therefore, we
only present results comparing CCM solutions based on the N\'{e}el,
spiral, and row-striped states as model states.

Thus, firstly, in Fig.\ \ref{E} we show our CCM results for the GS
energy per spin, $e\equiv E/N$, based on each of these states as model
states.  We show both raw LSUB$n$ results and the corresponding extrapolated
LSUB$\infty$ results using the extrapolation formula of Eq.\
(\ref{E_extrapo}).  It is clear that the LSUB$n$ results based on each
model state converge extremely rapidly as $n \rightarrow \infty$.
Figure \ref{E}(a) shows that there is no real evidence for a
discontinuity in slope at the critical values $\kappa_{c_{1}}$ (which
themselves depend only weakly upon the order $n$ of the LSUB$n$
approximation used).  If present at all, either in the raw LSUB$n$
results or the LSUB$\infty$ extrapolation, any such discontinuity can
only be very weak indeed.  

The overall accuracy of our results may also be ascertained by
examining the special case of $\kappa=0$ (corresponding to the
N\'{e}el order of the square-lattice HAF) and $\kappa=1$
(corresponding to the spiral, $\phi=\frac{1}{3}\pi$, order of the
triangular-lattice HAF).  Thus, for $\kappa=0$, the extrapolated GS
energy per spin, using our LSUB$n$ results with $n=\{4,6,8,10\}$ is
$e(\kappa=0) \approx -0.6697$, which may be compared with the results
$e(\kappa=0)=-0.6693(1)$ from a linked-cluster series expansion (SE)
technique,\cite{Zh:1991} and $e(\kappa=0)=-0.669437(5)$ from a
large-scale quantum Monte Carlo (QMC) simulation,\cite{Sa:1997} free
of the usual ``minus-sign-problems'' for this unfrustrated limiting
case where the Marshall-Peierls sign rules\cite{Ma:1955} applies.
Similarly, for $\kappa=1$, the extrapolated value using our LSUB$n$
results with $n=\{2,4,6,8\}$ is $e(\kappa=1) \approx -0.5511$, which
may again be compared with the results $e(\kappa=1)=-0.5502(4)$ from a
SE calculation,\cite{Zheng:2006} and $e(\kappa=1)=-0.5458(1)$ from a
QMC simulation.\cite{Ca:1999} It is worth noting that, unlike for the
square-lattice HAF, the nodal structure of the exact GS wave function
for the triangular-lattice HAF is unknown.  Hence, in this latter
$\kappa=1$ case the QMC minus-sign problem cannot be avoided.
Instead, a fixed-node approximation was made in Ref.\
\onlinecite{Ca:1999}, which was then relaxed using a stochastic
reconfiguration technique.  The resulting approximate result for the
GS is now only an upper bound, and our own CCM result is almost
certainly more accurate in this case.

In both Figs.\ \ref{E}(a) and \ref{E}(b) the extrapolated results are
shown only for those values of $\kappa$ in each case for which real
solutions exist for all of the LSUB$n$ approximations used in the
extrapolation.  In the case of the spiral state used as CCM model
states, shown in Fig.\ \ref{E}(a), the extrapolation is thus shown for
the range $0 < \kappa \lesssim 2.2$, since the LSUB8 solution shows an
instability for $\kappa \gtrsim 2.2$ as discussed previously.
Similarly, in Fig. \ref{E}(b), we can see that each of the LSUB$n$
results with $n > 2$ based on the row-striped state as model state,
terminate at some lower value, $\kappa_{t}(n)$, which itself depends
on the level $n$ of approximation.  Thus, we find $\kappa_{t}(4)
\approx 0.95 \pm 0.05$, $\kappa_{t}(6) \approx 1.15 \pm 0.05$,
$\kappa_{t}(8) \approx 1.35 \pm 0.05$, and $\kappa_{t}(10) \approx
1.35 \pm 0.05$.

As we have noted above, such termination points of our CCM LSUB$n$
calculations are always reflections of the true quantum phase
transitions in the system.\cite{Fa:2004,Bi:2009_SqTriangle,UJack_ccm}
They arise due to the solution to the CCM LSUB$n$ equations becoming
complex there, such that beyond any such termination point two
unphysical branches of complex conjugate solutions exist.  Conversely, in the
region where the true physical solution is real, there actually exists
another (unstable) real solution, which is both unphysical and
generally very difficult to determine numerically.  The physical
branch is also usually easily identifiable in practice as the one
which becomes exact in some limiting case.  At a termination point the
two real branches of solution (physical and unphysical) meet, and then
diverge again as wholly unphysical complex conjugate pairs beyond the
termination point for real solutions.  The values of $\kappa_{t}(n)$
may themselves be used to estimate the corresponding quantum critical
point (QCP) $\kappa_{c} = {\rm lim}_{n \rightarrow
  \infty}\kappa_{t}(n)$.  We do not attempt to do so here since it is
difficult to obtain values of $\kappa_{t}(n)$ with high precision,
since the number of iterations required to solve the CCM LSUB$n$
equations to a given level of accuracy generally increases
significantly as $\kappa \rightarrow \kappa_{t}(n)$.  Furthermore, we
have other more accurate criteria for determining the QCPs, as we discuss
below.  Nevertheless, even the crude values for $\kappa_{t}(n)$ above
for $n=\{4,6,8,10\}$ indicate that the row-striped state itself can
only be a possible candidate for the GS phase of our system for values
of $\kappa \gtrsim 1.5$.

\begin{figure}[t]
  \includegraphics[width=6cm,angle=270]{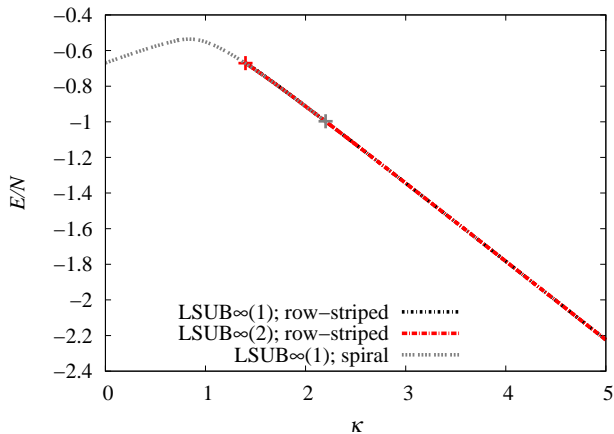}
  \caption{Extrapolated CCM LSUB$\infty$ results for the GS energy per
    spin, $E/N$, versus the frustration parameter $\kappa \equiv
    J_{2}/J_{1}$, for the spin-1/2 $J_{1}$--$J_{2}$ chevron-square
    lattice model (with $J_{1}>0$).  We show results based on the
    N\'{e}el and spiral model states for $\kappa < 2.2$, and on the
    row-striped model state for $\kappa > 1.4$, where these two
    (LSUB8) termination points (see text) are marked by $+$ 
    symbols.  The LSUB$\infty$(1) and LSUB$\infty$(2) curves are
    extrapolations using Eq.\ (\ref{E_extrapo}) and LSUB$n$ results
    with $n=\{2,4,6,8\}$, and $n=\{4,6,8,10\}$ respectively.}
\label{E_extrapolation}
\end{figure}

We show in Fig.\ \ref{E_extrapolation} our CCM LSUB$\infty$
extrapolated results for the GS energy per spin, $e \equiv E/N$, based
on the N\'{e}el, spiral and row-striped model states, using Eq.\
(\ref{E_extrapo}) and our corresponding LSUB$n$ results for the two
data sets $n=\{2,4,6,8\}$ and $n=\{4,6,8,10\}$.
We note in passing that if our LSUB$n$ results are instead fitted to a
form $e(n)=e_{0}+e_{1}n^{-\nu}$, the fitted value for the exponent
$\nu$ turns out to be very close to 2 for all values of $\kappa$
except those close to the termination points of the data set used, for
each of our model states.  The use of Eq.\ (\ref{E_extrapo}) is
thereby again justified for this model.  We see firstly from Fig.\
\ref{E_extrapolation} that the results are almost identical using both
data sets for the row-striped case illustrated.  Similar insensitivity
of our extrapolated results for the GS energy against the LSUB$n$ data
set used holds for the other model states too.  Secondly, we note how
our extrapolated results based on both the spiral and row-striped
states are very close to one another, although {\it not} identical, as
we discuss further below.  Thirdly, we note that in the $\kappa
\rightarrow \infty$ (decoupled spin-1/2 1D HAF chevron chains) limit,
the results shown in Fig.\ \ref{E_extrapolation} for the row-striped
state yield an extrapolated value of the energy per spin which
approaches the value $E/N=-0.4431\kappa$, which agrees precisely with
the exact result, $E/N=(\frac{1}{4}-\ln 2)\kappa$, from the Bethe ansatz
solution.\cite{Bethe:1931,Hulthen:1938}

\begin{figure}[t]
\includegraphics[angle=270,width=8.5cm]{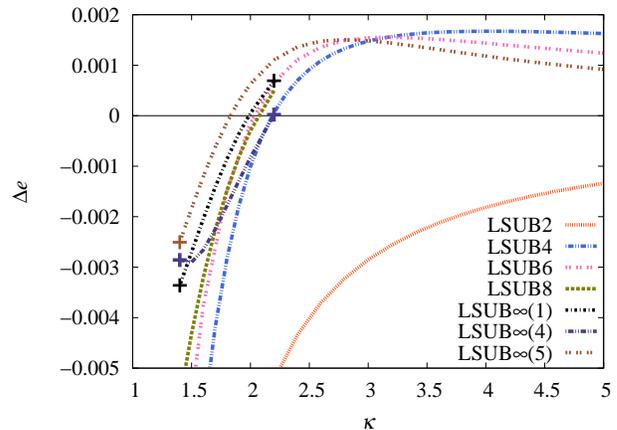}
\caption{(Color online) The GS energy difference, $\Delta e^{{\rm sr}}
  \equiv e^{{\rm s}} - e^{{\rm r}}$, with $e \equiv E/N$, between the
  spiral (s) and row-striped (r) states versus the frustration parameter
  $\kappa \equiv J_{2}/J_{1}$, for the spin-1/2 chevron-square lattice
  model (with $J_{1}>0$).  Results are shown using separate
  LSUB$n$ calculations for states s and r, with $n=\{2,4,6,8\}$, and
  also from separate LSUB$\infty(k)$ extrapolations for the states s and r
  using Eq.\ (\ref{E_extrapo}) and LSUB$n$ data sets separately: $k=1$, $n=\{2,4,6,8\}$; $k=4$, $n=\{4,6,8\}$; and $k=5$, $n=\{2,4,6\}$.  The respective
  termination points for the states are shown by $+$ symbols.}
\label{E_diff_spiral_rowStriped}
\end{figure}

In Fig.\ \ref{E_diff_spiral_rowStriped} we show directly the GS energy
difference, $e \equiv E/N$, between the spiral and row-striped states.
We display both raw results at various LSUB$n$ levels of approximation
and extrapolated results where we use Eq.\ (\ref{E_extrapo}) with
various LSUB$n$ data sets for each state separately before taking the
difference.  In each case the extrapolated results are shown for
values between the respective termination points for each of the two
states.  Although the energy difference is small, it is well within
our level of accuracy.

\begin{figure*}[t]
\mbox{
\subfigure[]{\scalebox{0.3}{\includegraphics[angle=270]{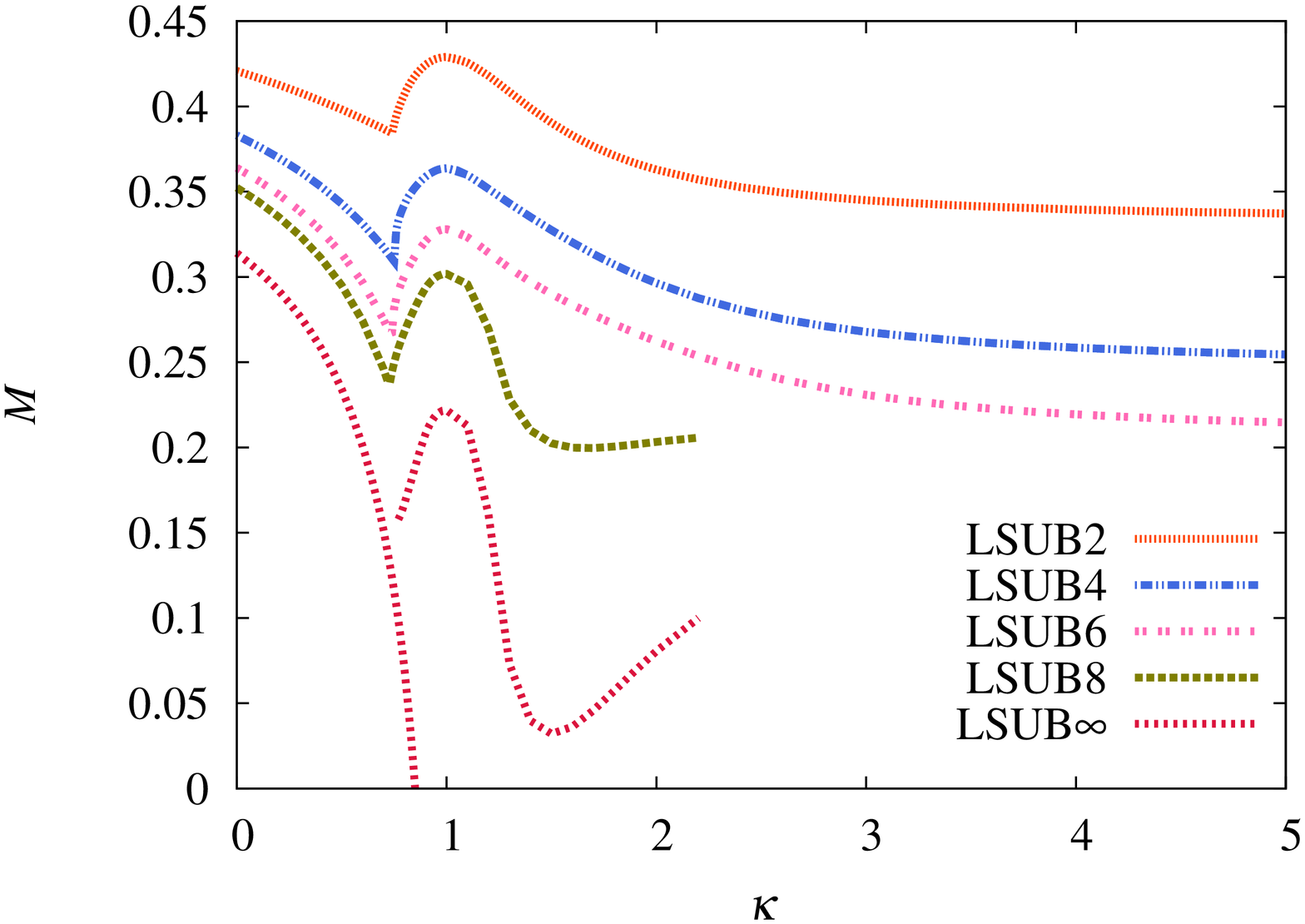}}}
\subfigure[]{\scalebox{0.3}{\includegraphics[angle=270]{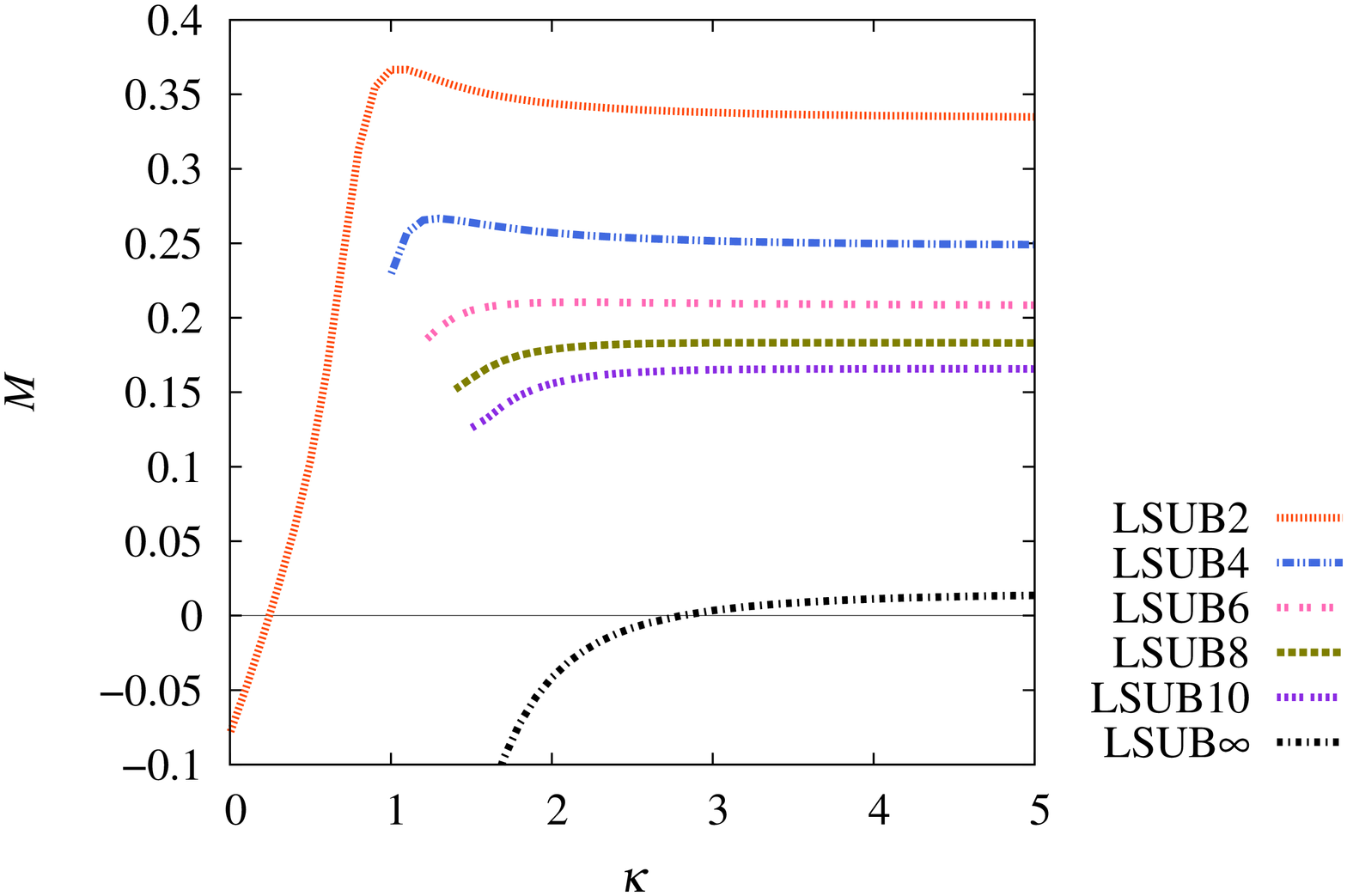}}}
}
\caption{(Color online) CCM LSUB$n$ results for the GS magnetic order
  parameter, $M$, versus the frustration parameter $\kappa \equiv
  J_2/J_{1}$, for the spin-1/2 $J_{1}$--$J_{2}$ chevron-square lattice
  model (with $J_{1}>0$), using (a) the N\'{e}el and spiral states of
  Fig.\ \ref{model_bonds}(a) and \ref{model_bonds}(b) respectively as
  model states with $n=\{2,4,6,8\}$, and (b) the row-striped state of Fig.\ \ref{model_bonds}(c) as model state with $n=\{2,4,6,8,10\}$.  In
  each case we show the extrapolated LSUB$\infty$ result using the data
  sets shown and the respective extrapolation formula of
  Eq.~(\ref{M_extrapo_standard}) and (\ref{M_extrapo_frustrated})
  respectively.}
\label{M}
\end{figure*}

It is interesting to note that while the spiral state has lower energy
than the row-striped state for all values ($\kappa < 5$) of the
frustration parameter shown at the lowest LSUB2 level of
approximation, there is a crossover point above which the row-striped
state lies lower in energy than the spiral state for each of the
higher LSUB4 and LSUB6 levels of approximation.  This is also true at
the LSUB8 level, but the crossover point is now rather close to the
point ($\kappa \approx 2.2$) at which the corresponding LSUB8 solution
for the spiral state becomes unstable.  The crossover point for the
extrapolated results clearly shows some (relatively small) sensitivity
with respect to which set of LSUB$n$ results is used to perform the
extrapolation.  In view of the instability of the spiral (LSUB8)
results in the relevant range, our best estimate undoubtedly comes
from using the data set $n=\{2,4,6\}$.  This gives an energy crossover
at $\kappa \approx 1.83$, above which the spiral state ceases to be
the lowest-energy solution.  This value is itself in remarkably good
agreement with our estimate above for the (termination) point at
$\kappa \approx 1.5$, above which the row-striped state is itself
stable.  Indications so far thus provide reasonably compelling
evidence for a second QCP at a value $\kappa_{c_{2}} \approx 1.7(3)$,
above which the spiral state ceases to be the stable GS.

In order to shed more light on this transition we now present our
corresponding CCM results for the magnetic order parameter, $M$,
defined to be the average local on-site magnetization.  In Fig.\
\ref{M}(a) we show LSUB$n$ results for the N\'{e}el and spiral states
as CCM model states with $n \leq 8$.
Note that for the N\'{e}el state we also have LSUB10 results, not
shown on the figure since we do not have corresponding LSUB10 results
for the spiral state.  Once again, it is clear that the LSUB8
results for the spiral state behave rather differently than the
LSUB$n$ results with $n \leq 6$ in the region $\kappa > 1$ as the
instability around $\kappa \approx 2.2$ is approached.  We also show
in Fig.\ \ref{M}(a) our extrapolated LSUB$\infty$ result using Eq.\
(\ref{M_extrapo_standard}) and the data set $n=\{2,4,6,8\}$.  Once
again, the overall accuracy of our results may be gauged by examining
the two special cases of the pure isotropic HAF with NN interactions
only on the square lattice ($\kappa = 0$) and the triangular lattice
($\kappa = 1$).

For the square lattice we find $M(\kappa=0)=0.3087$ using Eq.\
(\ref{M_extrapo_standard}) with LSUB$n$ results for $n=\{4,6,8\}$ and
$M(\kappa=0)=0.3074$ from $n=\{4,6,8,10\}$.  These values are again in
excellent agreement with the corresponding results
$M(\kappa=0)=0.3070(3)$ from a large-scale QMC
simulation,\cite{Sa:1997} and $M(\kappa=0)=0.307(1)$ from a
linked-cluster SE calculation,\cite{Zh:1991} Similarly, for the
triangular lattice we find $M(\kappa=1)=0.2219$ using Eq.\
(\ref{M_extrapo_standard}) with LSUB$n$ results for $n=\{2,4,6,8\}$
and $M(\kappa=1)=0.1893$ from $n=\{4,6,8\}$.  Again, we are in
excellent agreement with other values $M(\kappa=1)=0.205(10)$ from a
fixed-node QMC simulation,\cite{Ca:1999}, and $M(\kappa=1)=0.19(2)$
from a linked-cluster SE calculation.\cite{Zheng:2006}

Figure \ref{M}(a) is completely consistent with the result that the
magnetic order parameter $M \rightarrow 0$ at the critical point
$\kappa = \kappa_{c_{1}}$ where the collinear N\'{e}el order for
$\kappa < \kappa_{c_{1}}$ yields to the noncollinear spiral order.
Our best estimate for $\kappa_{c_{1}}$, based on the results for $M$,
is $\kappa_{c_{1}}=0.72(1)$, where the error is estimated from a
sensitivity analysis with respect to which LSUB$n$ results are used in
the extrapolation.  There is also definite evidence from Fig.\
\ref{M}(a) that in the spiral regime ($\kappa > \kappa_{c_{1}}$), the
spiral order parameter $M$ again becomes zero (or very close to zero)
at a second critical value, $\kappa_{c_{2}} \approx 1.4$, which we
discuss in more detail later.

\begin{figure}[!b]
\includegraphics[angle=270,width=8.5cm]{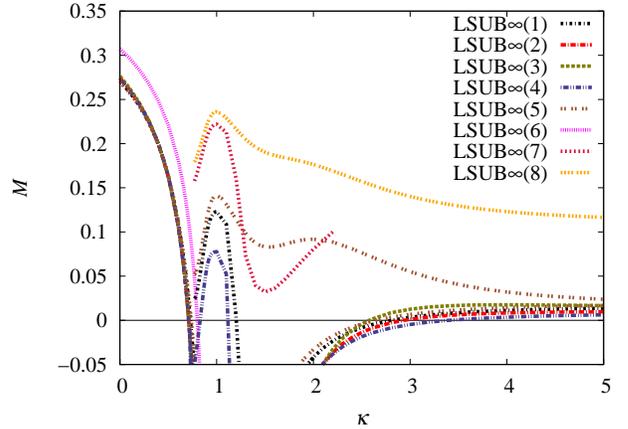}
\caption{(Color online) Various extrapolated CCM results for the GS
  magnetic order parameter, $M$, versus the frustration parameter
  $\kappa \equiv J_2/J_1$, based on the N\'{e}el (left curves) and
  spiral (central curves) model states and the row-striped (right
  curves) model state, for the spin-1/2 $J_{1}$--$J_{2}$
  chevron-square lattice model (with $J_{1}>0$).  The curves
  LSUB$\infty(k)$ with $k=1,2,\cdots,5$ use the extrapolation scheme
  of Eq.~(\ref{M_extrapo_frustrated}) and LSUB$n$ data sets
  respectively: $k=1$, $n=\{2,4,6,8\}$; $k=2$, $n=\{4,6,8,10\}$;
  $k=3$, $n=\{6,8,10\}$, $k=4$, $n=\{4,6,8\}$, and $k=5$,
  $n=\{2,4,6\}$.  The curves LSUB$\infty(k)$ with $k=6,7,8$ use the
  extrapolation scheme of Eq.~(\ref{M_extrapo_standard}) and LSUB$n$
  data sets, respectively: $k=6$, $n=\{4,6,8,10\}$, for the N\'{e}el
  model state; and $k=7$, $n=\{2,4,6,8\}$, and $k=8$, $n=\{2,4,6\}$ for the spiral model state.}
\label{M_extrapolation}
\end{figure}

Corresponding CCM LSUB$n$ results for the magnetic order parameter,
$M$, for the row-striped state are shown in Fig.\ \ref{M}(b) with
$n=\{2,4,6,8,10\}$.  In this highly frustrated regime, if we fit our
results to a form $M(n)=m_{0}+m_{1}n^{-\nu}$, as in Eq.\ (\ref{M_extrapo_nu}), we find that the fitted
value of the exponent $\nu$ turns out to be very close to 0.5 over
almost the entire range of values of $\kappa$, independent of which
LSUB$n$ data set is used for the fit.  Hence, in this case it is
appropriate to use the extrapolation scheme of
Eq.~(\ref{M_extrapo_frustrated}), and in Fig.\ \ref{M}(b) we use the
data set $n=\{2,4,6,8\}$ for illustration purposes and compatibility
  with Fig.\ \ref{M}(a).  What is most striking about the extrapolated
  result is that $M$ appears to be essentially zero (or less than
  zero, and hence unphysical) over the entire regime in which the
  row-striped state exists as a candidate for the stable GS phase.  In
  Fig.\ \ref{M_extrapolation} we present various extrapolations of our
  CCM LSUB$n$ data, from which we observe their relative insensitivity
  to the choice of data set.

\begin{figure*}[!t]
\subfigure[]{\scalebox{0.4}{\includegraphics{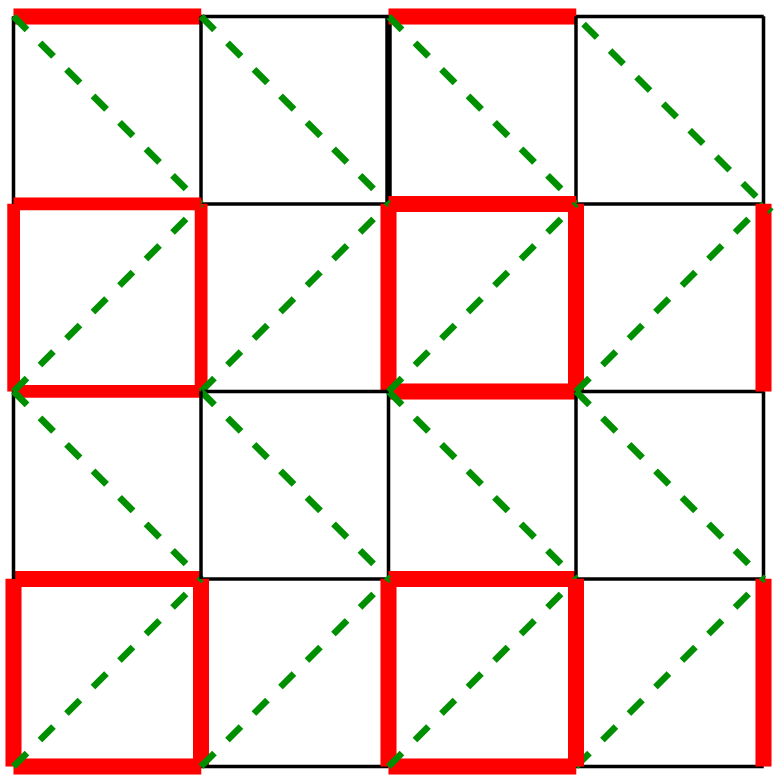}}} \quad
\subfigure[]{\scalebox{0.4}{\includegraphics{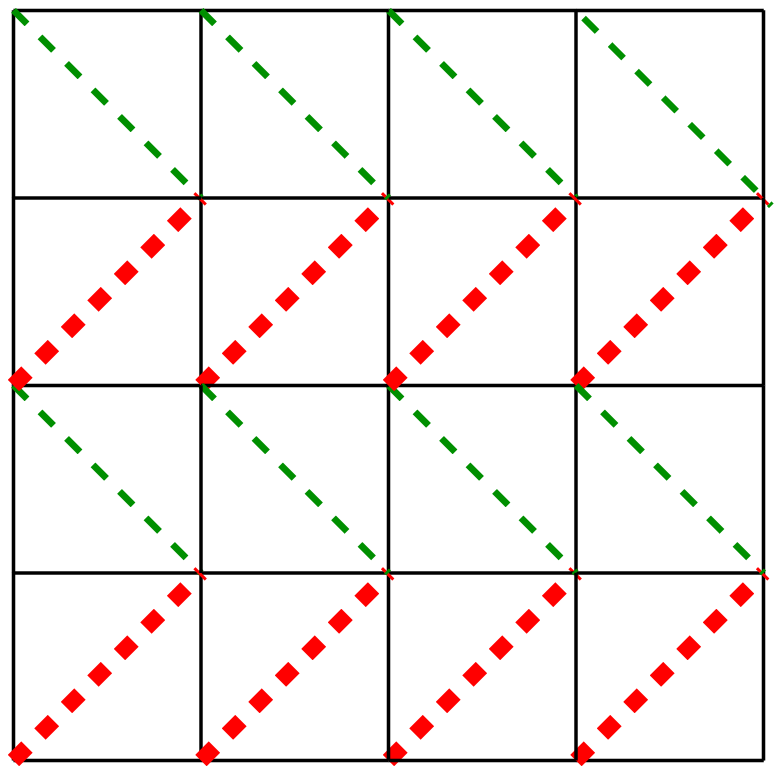}}} \quad
\subfigure[]{\scalebox{0.4}{\includegraphics{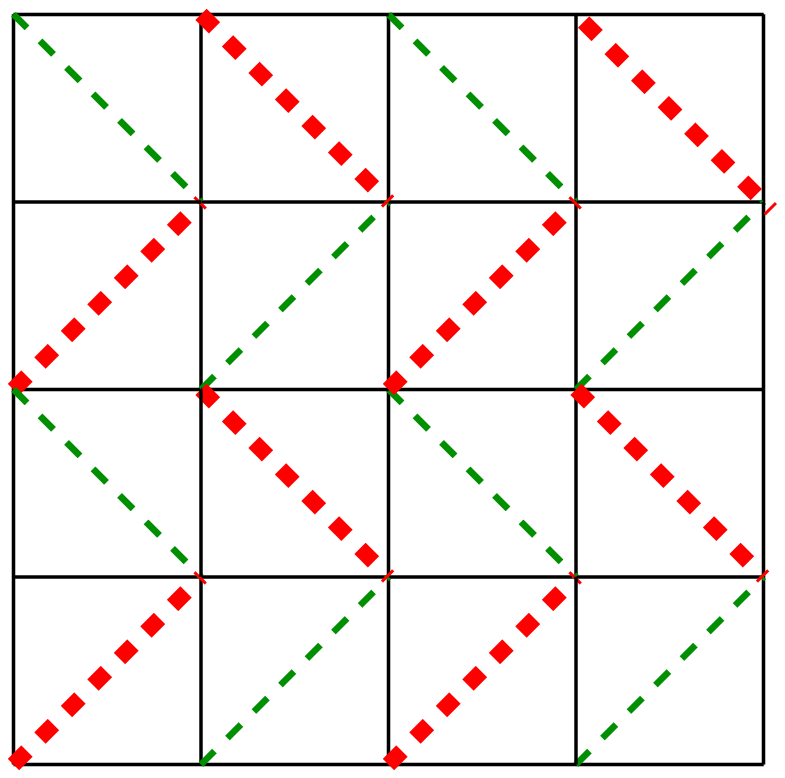}}} 
\caption{(Color online) The perturbations (fields), $F=\delta\,
  \hat{O}$, for the various forms of valence-bond crystal (VBC)
  susceptibility considered, namely: (a) plaquette (PVBC), $\chi_{p}$;
  (b) parallel-dimer ($\parallel$DVBC), $\chi_{{\rm d}_{\parallel}}$;
  and (c) perpendicular-dimer ($\perp$DVBC), $\chi_{{\rm d}_{\perp}}$.
  In case (a) the thick (red) and thin (black) solid lines correspond
  respectively to strengthened and weakened $J_{1}$ exchange
  couplings, where $\hat{O}_{\rm p} = \sum_{\langle i,j \rangle}
  a_{ij} \mathbf{s}_{i}\cdot\mathbf{s}_{j}$, and the sum runs over all
  NN bonds of the square lattice, with $a_{ij}=+1$ and $-1$ for thick
  (red) and thin (black) solid lines respectively, as shown.  In cases (b)
  and (c) the thick (red) and thin (green) dashed lines
  correspond respectively to strengthened and weakened $J_{2}$
  exchange couplings, and $\hat{O}_{{\rm d}} =
  \sum_{\langle \langle i,k \rangle \rangle} a_{ik}
  \mathbf{s}_{i}\cdot\mathbf{s}_{k}$, and the sum runs over the
  diagonal bonds of the chevron-square lattice, with $a_{ik}=+1$ and
  $-1$ for thick (red) and thin (green) dashed lines
  respectively, as shown in each case.}
\label{X_patterns}
\end{figure*}

All of our results so far are consistent with the interpretation that
the spiral state exists as the stable GS phase only in a regime
$\kappa_{c_{1}} < \kappa < \kappa_{c_{2}}$, where $\kappa_{c_{2}}
\approx 1.7(5)$.  However, the estimates so far for $\kappa_{c_{2}}$
are based on the energy crossing point between the spiral and
row-striped states, the termination points for the LSUB$n$ row-striped
results, and the magnetic order parameter results for the row-striped
state.  Nevertheless, the latter results also indicate that, since $M
\approx 0$ everywhere that the row-striped solution exists as a real
solution, the row-striped state itself is actually never realized as
the stable GS phase.  

In order to obtain more information about the stable GS phase that the
system actually enters after the melting of spiral order at
$\kappa_{c_{2}}$ we now investigate its susceptibility to various
forms of valence-bond crystalline (VBC) orders.  In particular, we consider
the plaquette valence-bond crystalline (PVBC) order illustrated in Fig.\
\ref{X_patterns}(a) and two forms of dimer valence-bond crystalline (DVBC)
order illustrated in Figs.\ \ref{X_patterns}(b) and (c), respectively,
which we refer to henceforth as parallel ($\parallel$DVBC) and
perpendicular ($\perp$DVBC) forms.
In each case we consider the response of the system to a corresponding
field operator, $F=\delta\, \hat{O}$, which is added to the Hamiltonian
of Eq.\ (\ref{H}) (and see, e.g., Ref.\
\onlinecite{Darradi:2008_J1J2mod}).  The corresponding operators
$\hat{O}_{\rm p}$, $\hat{O}_{{\rm d}_{\parallel}}$, and $\hat{O}_{{\rm
    d}_{\perp}}$ for PVBC, $\parallel$DVBC, and $\perp$DVBC order are
shown graphically in Fig.\ \ref{X_patterns} and defined explicitly in
the caption.  For each form of VBC order considered, we then calculate
the perturbed energy per site, $e(\delta) \equiv E(\delta)/N$, for the
modified Hamiltonian $H + F$, at various LSUB$n$ levels of
approximation, using the N\'{e}el and row-striped states as CCM model
states.  The corresponding susceptibility, 
\begin{equation}
\chi \equiv -\left. (\partial^2{e(\delta)})/(\partial {\delta}^2)   \label{Eq_X}
\right|_{\delta=0}\,,
\end{equation}
is then calculated.  Clearly the phase
corresponding to the model state used becomes unstable against the
specified form of VBC order when the corresponding extrapolated
inverse susceptibility, $\chi^{-1}_{d}$, goes to zero.

The most direct and most unbiased way to extrapolate the results is to
use the LSUB$n$ scheme for the perturbed energy
\begin{equation}
e^{(n)}(\delta) =
e_{0}(\delta)+e_{1}(\delta)n^{-\nu},   \label{Extrapo_dBonds}
\end{equation}
where the exponent $\nu$ is also a fitting parameter.  Generally, as
we discuss below, the fitted value of $\nu$ is close to the value 2,
as expected from our standard energy extrapolation scheme of Eq.\
(\ref{E_extrapo}), over most of the range of values of $\kappa$
pertaining to the model state used, except very near any corresponding
termination point for the given state, where $\nu$ then falls sharply.
Alternatively, we have also found previously\cite{Farnell:2011} that
our LSUB$n$ values $\chi(n)$ may often themselves be directly
extrapolated very accurately using the same scheme as in Eq.\
(\ref{E_extrapo}) for the GS energy, viz., $\chi(n) =
d_{0}+d_{1}n^{-2}+d_{2}n^{-4}$, at least in regions not close to a
divergence of $\chi$.  We saw similarly\cite{Farnell:2011} that a
corresponding extrapolation of the inverse susceptibility,
\begin{equation}
\chi^{-1}(n) = x_{0}+x_{1}n^{-2}+x_{2}n^{-4}\,,             \label{Extrapo_inv-chi-1}
\end{equation}
also often gives consistent results that agree closely with those from
the corresponding above extrapolation of $\chi$ itself, except again
in regions close to where $\chi^{-1} \rightarrow 0$.  However, since
we shall specifically be interested in some cases here in precisely such
regions over an extended range of values of the frustration parameter
$\kappa$, we also use the fitting function
\begin{equation}
\chi^{-1}(n) = y_{0}+y_{1}n^{-\nu}\,,            \label{Extrapo_inv-chi-2}
\end{equation}
with an exponent $\nu$ that is itself a fitting parameter, on
appropriate occasions.

\begin{figure}[t]
  \includegraphics[width=6cm,angle=270]{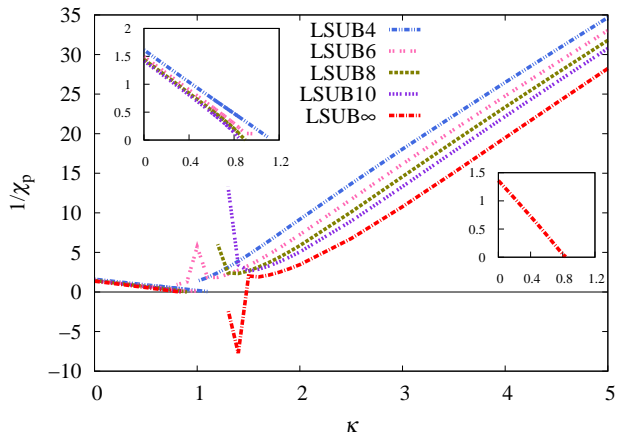}
  \caption{(Color online) CCM LSUB$n$ results for the inverse
    plaquette susceptibility, $1/\chi_{{\rm p}}$, appropriate to the
    PVBC ordering of Fig.~\ref{X_patterns}(a), versus the frustration
    parameter $\kappa \equiv J_{2}/J_{1}$, using the N\'{e}el
    (left curves) and row-striped (right curves) states as model
    states for the spin-1/2 $J_{1}$--$J_{2}$ Heisenberg
    antiferromagnet on the chevron-square lattice (with $J_{1}>0$).
    For both model states we show the LSUB$n$ results with
    $n=\{4,6,8,10\}$, plus the corresponding extrapolated results
    LSUB$\infty$ using this data set with the extrapolation scheme of
    Eq.~(\ref{Extrapo_dBonds}).}
\label{plaqXcpty}
\end{figure}

We first present results in Fig.\ \ref{plaqXcpty} for the inverse
plaquette susceptibility, $\chi^{-1}_{{\rm p}}$, pertaining to the
PVBC ordering illustrated in Fig.\ \ref{X_patterns}(a).
We show explicitly the LSUB$n$ results with $n=\{4,6,8,10\}$ based on
both the N\'{e}el and row-striped states as CCM model states, together
with the corresponding LSUB$\infty$ extrapolations using Eq.\
(\ref{Extrapo_dBonds}) with these data sets.  As noted above, the
fitted value of the exponent $\nu$ in Eq.\ (\ref{Extrapo_dBonds}) is
very close to 2 except in very narrow regions close to
$\kappa_{c_{1}}$ on the N\'{e}el side and close to $\kappa_{c_{2}}$ on
the row-striped side.  It is interesting to note that the N\'{e}el
phase has a high susceptibility to PVBC ordering, although
$\chi^{-1}_{{\rm p}} \rightarrow 0$ only at a single value which is
very close to our previous best estimate for $\kappa_{c_{1}}$.  While
the row-striped state, on the other hand, is clearly much less
susceptible to PVBC ordering, the corresponding LSUB$\infty$
estimate for $\chi^{-1}_{{\rm p}}$ again becomes very close to zero at
a value $\kappa_{c_{2}} \approx 1.4$, which is again compatible with
our previous estimates for $\kappa_{c_{2}}$.  Presumably, if we were
also to perform calculations for $\chi^{-1}_{{\rm p}}$ for the spiral
phase, it too would be nonzero (and positive) everywhere except at the
values $\kappa_{c_{1}}$ and $\kappa_{c_{2}}$.  However, such
calculations for the spiral phase are computationally very expensive,
and as noted previously can be performed only for relatively lower
LSUB$n$ orders of approximation.  Even for the collinear row-striped
state the number $N_{f}$ of fundamental spin configurations included
in our CCM calculation at the LSUB10 level is 2160176.

While the results shown in Fig.\ \ref{plaqXcpty} corroborate our
previous estimates for the QCPs at $\kappa_{c_{1}}$ (at which N\'{e}el
order yields to spiral order) and $\kappa_{c_{2}}$ (at which spiral
order ceases), they essentially add nothing new.  Thus, it is clear
that the PVBC state cannot be the stable GS phase for any range of
values $\kappa > \kappa_{c_{2}}$, since $\chi^{-1}_{{\rm p}}$ does not
vanish over any finite range in this regime.  The fact that
$\chi^{-1}_{{\rm p}}$ vanishes at individual points (viz.,
$\kappa_{c_{1}}$ and $\kappa_{c_{2}}$ here) is simply a reflection of
these being actual QCPs.  Thus, at a QCP we expect that the system
will indeed become infinitely susceptible to {\it all} forms of
ordering (or, at least, those which are compatible with the symmetries
of the actual and trial states).

We now turn our attention to the two DVBC states of Figs.\
\ref{X_patterns}(b) and (c) as possible GS phases for $\kappa >
\kappa_{c_{2}}$.  Firstly, since our results for the inverse
susceptibility, $1/\chi_{{\rm d}_{\perp}}$, for $\perp$DVBC ordering
are qualitatively very similar to those shown in Fig.\ \ref{plaqXcpty}
for PVBC ordering, we do not show them.  Again, we find that using the
N\'{e}el model state $\chi^{-1}_{{\rm d}_{\perp}}$ vanishes at a
single point, which is itself completely compatible with our previous
estimates for $\kappa_{c_{1}}$, and using the row-striped model state
$\chi^{-1}_{{\rm d}_{\perp}}$ vanishes at a single point compatible
with our previous estimates for $\kappa_{c_{2}}$.  Thus, again, the
$\perp$DVBC state is excluded as a stable GS phase in any range
$\kappa > \kappa_{c_{2}}$.

\begin{figure*}[!t]
\mbox{
\subfigure[]{\scalebox{0.35}{\includegraphics[angle=270]{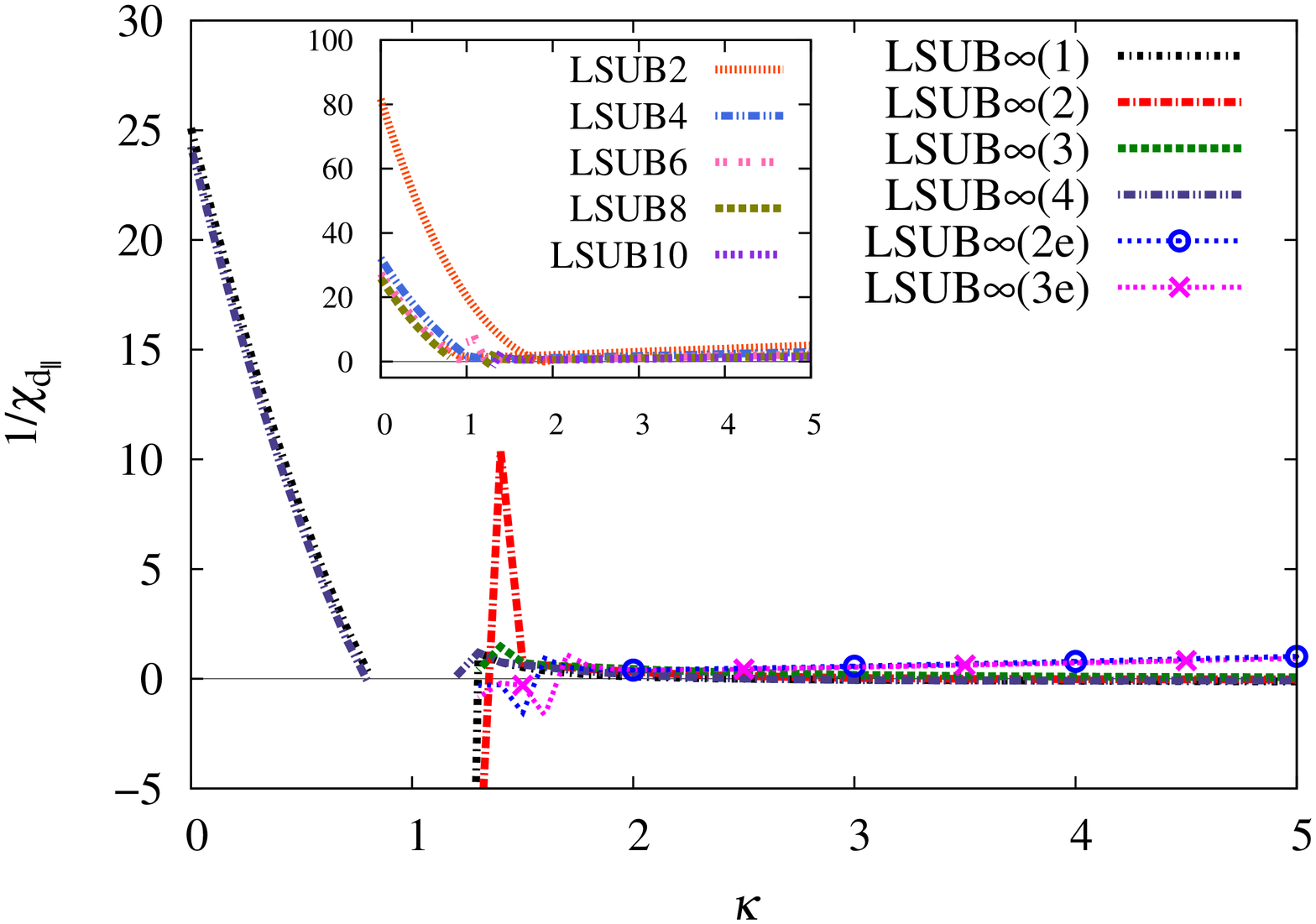}}}
\subfigure[]{\scalebox{0.35}{\includegraphics[angle=270]{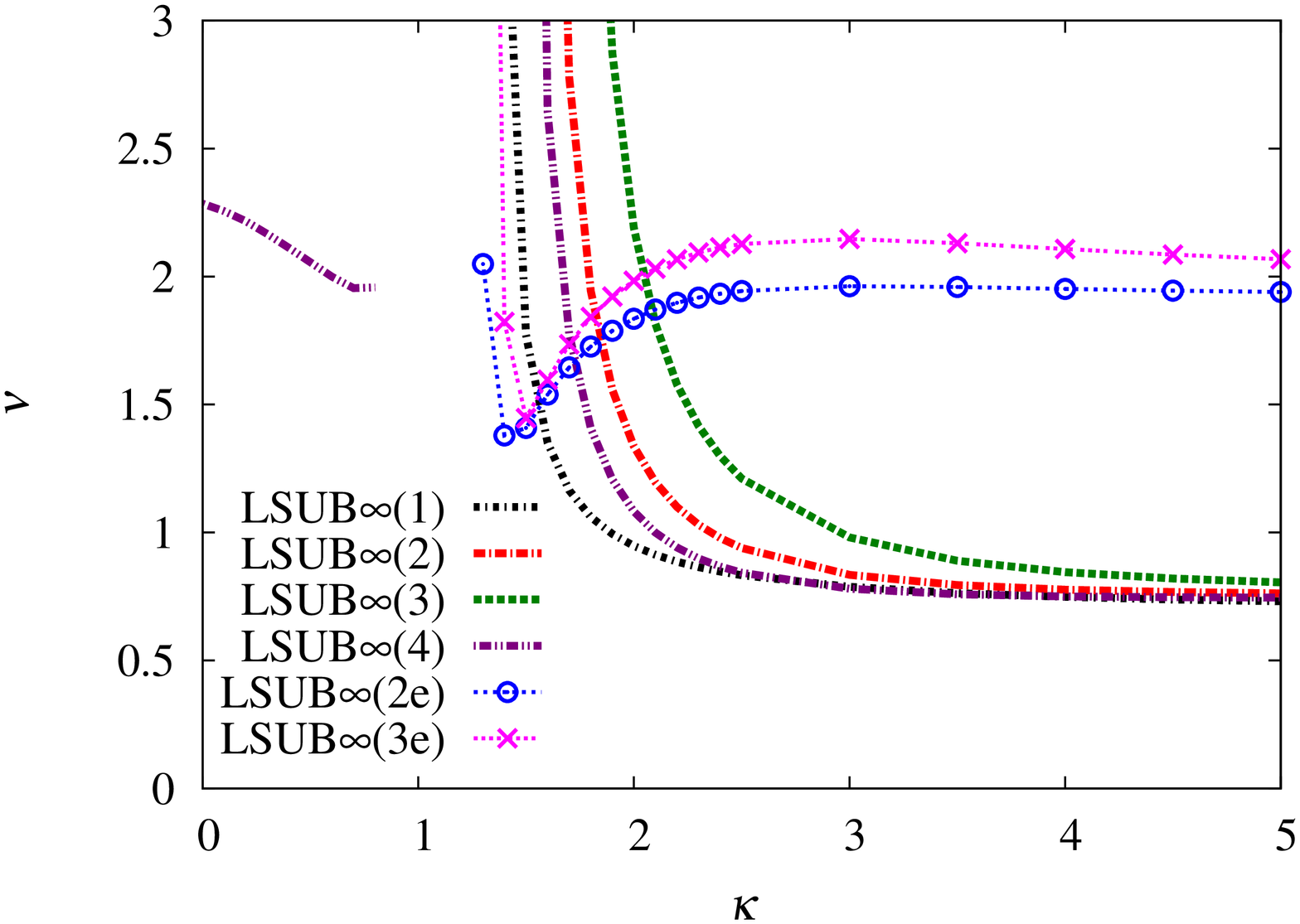}}}
}
\caption{(Color online) (a) CCM results for the inverse parallel-dimer 
  susceptibility, $1/\chi_{d_{\parallel}}$, appropriate to the
  $\parallel$DVBC ordering of Fig.~\ref{X_patterns}(b), versus the
  frustration parameter $\kappa \equiv J_{2}/J_{1}$, using the
  N\'{e}el (left curves) and row-striped (right curves) states of
  Figs.\ \ref{model_bonds}(a) and \ref{model_bonds}(c), respectively,
  as model states for the spin-1/2 $J_{1}$--$J_{2}$ Heisenberg
  antiferromagnet on the chevron-square lattice ($J_{1}>0$).  For both
  model states we show the LSUB$n$ results with $n=\{2,4,6,8,10\}$ in
  the inset, while the main figure shows various LSUB$\infty$
  extrapolations.  The curves LSUB$\infty(k)$ (without symbols attached) with
  $k=1,\cdots,4$ use the extrapolation scheme of Eq.\
  (\ref{Extrapo_inv-chi-2}), while the curves LSUB$\infty(k{\rm e})$
  (with symbols attached) with $k=2,3$ use the extrapolation scheme
  of Eq.~(\ref{Extrapo_dBonds}), in both cases using the respectively
  LSUB$n$ data sets: $k=1$, $n=\{2,4,6,8\}$; $k=2$, $n=\{4,6,8,10\}$;
  $k=3$, $n=\{6,8,10\}$; and $k=4$, $n=\{4,6,8\}$.  (b) The respective
  fitted values of the exponent $\nu$ in the extrapolation schemes of
  Eqs.\ (\ref{Extrapo_dBonds}) and (\ref{Extrapo_inv-chi-2}), for the
  same fits as shown in subfigure (a) to the left.}
\label{dimer_parallel_Xcpty}
\end{figure*}

By contrast, our results for the inverse susceptibility, $1/\chi_{{\rm
    d}_{\parallel}}$, for $\parallel$DVBC ordering, shown in Fig.\
\ref{dimer_parallel_Xcpty}(a) are qualitatively quite different in the
regime $\kappa > \kappa_{c_{2}}$.  In the inset to Fig.\
\ref{dimer_parallel_Xcpty}(a) we show the raw LSUB$n$ results for
$n=\{2,4,6,8,10\}$ using both the N\'{e}el and row-striped states as
CCM model states, while in the main figure we show several LSUB$\infty$
extrapolations based on both Eqs.~(\ref{Extrapo_dBonds}) and
(\ref{Extrapo_inv-chi-2}), and using various LSUB$n$ data sets for the
fitting procedure.
In Fig.\ \ref{dimer_parallel_Xcpty}(b) we also show the fitted values
for the exponent $\nu$ in the corresponding extrapolation schemes.

In the first place we find that on the N\'{e}el side all of the
extrapolations yield almost identical results, with
$\chi^{-1}_{d_{\parallel}} \rightarrow 0$ at values $\kappa
\rightarrow \kappa_{c_{1}}$, which are in agreement with all previous
estimates.  Secondly, however, we now see from Fig.\
\ref{dimer_parallel_Xcpty}(a) that using the row-striped state as CCM
model state, all of our extrapolations yield the result that
$1/\chi_{{\rm d}_{\parallel}}$ is zero (or very close to zero) for
{\it all} values $\kappa > \kappa_{{c_{2}}} \approx 1.5$.  We thus
now have very strong evidence indeed that for $\kappa >
\kappa_{c_{2}}$, the GS spiral ordering present for $\kappa_{c_{1}} <
\kappa < \kappa_{c_{2}}$ gives way to $\parallel$DVBC ordering as the
stable GS configuration.

Figure \ref{dimer_parallel_Xcpty}(b) shows that when $\chi_{{\rm
    d}_{\parallel}}$ is calculated using Eq.\ (\ref{Extrapo_dBonds})
as the extrapolation scheme the fitted exponent $\nu$ is very close to
the expected value 2 for a very wide range of values of $\kappa$, as
in our standard GS energy scheme of Eq.\ (\ref{E_extrapo}), except for
values of $\kappa$ close to $\kappa_{c_{2}}$, where it drops.  In
practice we calculate the second derivative of $e(\delta)$ in Eq.\
(\ref{Eq_X}) numerically using values $\delta = 0\,,\pm d$, typically
with $d=0.001$, and the value of $\nu$ plotted in Fig.\
\ref{dimer_parallel_Xcpty}(b) is essentially identical for these three
values of $\delta$ used.  By contrast, when $1/\chi_{{\rm
    d}_{\parallel}}$ is calculated using Eq.\
(\ref{Extrapo_inv-chi-2}), the fitted exponent $\nu$ is seen to lie
very close to the value 0.75 for a very wide range of values of
$\kappa$, except for values close to $\kappa_{c_{2}}$ where it rises
rapidly.  Interestingly, this same value of 0.75 was also observed
previously for the spin-1/2 $J_{1}$--$J_{2}$ Heisenberg
antiferromagnet on the 2D checkerboard lattice (viz., the APP
model),\cite{Bishop:2012_checkerboard} when using a similar state as
CCM model state in a region where the order parameter for that state
was zero everywhere, and where a corresponding DVBC state also
provided the stable GS configuration.

\section{SUMMARY}
\label{summary_sec}
We have investigated the $T=0$ GS phase diagram of an $s=\frac{1}{2}$
$J_{1}$--$J_{2}$ HAF on a chevron-square lattice, the classical ($s
\rightarrow \infty$) version of which has only two GS phases, viz., a
N\'{e}el phase for $J_{2}/J_{1} \equiv \kappa < \kappa_{{\rm cl}} =
\frac{1}{2}$, and a spirally-ordered phase for $\kappa > \kappa_{{\rm
    cl}}$.  The model interpolates continuously between 2D HAFs on the
square ($\kappa = 0$) and triangular ($\kappa=1$) lattices, and also
extrapolates in the $\kappa \rightarrow \infty$ limit to uncoupled 1D
HAF chains.

We found, as expected, that the quantum fluctuations present in the
$s=\frac{1}{2}$ model tend to stabilize the collinear N\'{e}el-ordered
state, which is well established as the GS phase for the square-lattice HAF, to larger
values of the frustration parameter $\kappa$ than in the classical
case.  We found a first QCP at $\kappa = \kappa_{c_{1}} \approx
0.72(1)$ at which the N\'{e}el order gives way to spiral order.  The
pitch angle $\phi$ of the spiral phase (that measures the deviation
from N\'{e}el order corresponding to $\phi=0$) was found to increase
markedly more rapidly as $\kappa$ is increased beyond $\kappa_{c_{1}}$ than in the classical version as $\kappa$ is increased beyond
$\kappa_{{\rm cl}}$.

Although the transition at $\kappa=\kappa_{c_{1}}$ appears still to be
a continuous (second-order) one, just as for the classical model at
$\kappa = \kappa_{{\rm cl}}$, we cannot from our results rule out a
weakly first-order transition at which the pitch angle $\phi$
undergoes a discontinuous jump from a value of zero (N\'{e}el order)
for $\kappa < \kappa_{c_{1}}$ to a nonzero value (spiral order)
infinitesimally above $\kappa_{c_{1}}$.  The angle $\phi$ becomes
exactly $60^{\circ}$ at all levels of CCM LSUB$n$ approximation at
$\kappa=1$, corresponding to the exact three-sublattice relative
$120^{\circ}$ spin ordering that is well established as the GS phase
for the spin-1/2 triangular-lattice HAF.

At still higher values of the frustration parameter, $\kappa > 1$, we
found that the pitch angle $\phi$ of the $s=\frac{1}{2}$ model
approaches the limiting value of $90^{\circ}$ (which corresponds to
N\'{e}el AFM ordering along the 1D chevron chains of $J_{2}$ bonds)
much faster than in the classical ($s \rightarrow \infty$) model.  In
turn this led us to consider other possible collinear phases in this
regime, which never form the stable GS phase in the classical case,
except in the precise $\kappa \rightarrow \infty$ limit, but which
comprise an infinitely degenerate family of states corresponding to
N\'{e}el ordering along the 1D chevron chains.  We found that the {\it
  order by disorder} mechanism distinguishes one such state from the
rest, so that quantum fluctuations in the $s=\frac{1}{2}$ case favor
it to lie lowest in energy.  Furthermore, above some value of
the frustration parameter, $\kappa > \kappa_{c_{2}}$, this state was
seen to become lower in energy too than the spiral state for the
$s=\frac{1}{2}$ model.  However, when we then calculated its magnetic
order parameter, we found it to be zero, within very small numerical
error bars, everywhere.

We were thus led to the conclusion that the spiral order in the
$s=\frac{1}{2}$ model does not persist for all values $\kappa >
\kappa_{c_{1}}$, but only for a finite range $\kappa_{c_{1}} < \kappa <
\kappa_{c_{2}}$, unlike in the classical model where it persists for
{\it all} values $\kappa > \kappa_{{\rm cl}}$.  Instead, we found
again that the magnetic order parameter of the spiral phase seems to
vanish for values of $\kappa$ above a certain critical value, which is
itself consistent within small errors bars with the energy crossing
point of the spiral state with the nonclassical collinear state
mentioned.

Since this latter collinear state itself has a vanishing magnetic
order parameter in the same region, as already noted, it must itself
yield to a state with different ordering.  By comparison with other
related depleted $J_{1}$--$J_{2}$ models, we were led to consider that
this collinear state might be susceptible to some form of VBC
ordering, and we therefore calculated its susceptibility to both PVBC
ordering and two forms of DVBC ordering.  From among these we found
unequivocal evidence that there is a second QCP in the $s=\frac{1}{2}$
model at a value $\kappa=\kappa_{c_{2}} \approx 1.5(1)$ at which the
spiral order present for $\kappa_{c_{1}} < \kappa < \kappa_{c_{2}}$
melts and gives way for values $\kappa > \kappa_{c_{2}}$ to a
parallel-dimer VBC ($\parallel$DVBC) order.

In conclusion, the $J_{1}$--$J_{2}$ HAF on the chevron-square lattice
has been found to provide a rich GS phase diagram for the extreme
quantum case, $s=\frac{1}{2}$.  It would certainly be interesting,
therefore, both to confirm our results using other theoretical methods
and to examine the corresponding $s=1$ version of the model.

\section*{ACKNOWLEDGMENTS}
We thank the University of Minnesota Supercomputing Institute for
Digital Simulation and Advanced Computation for the grant of
supercomputing facilities, on which we relied heavily for the
numerical calculations reported here.


\begin{thebibliography}{99}

\bibitem{2D_magnetism_1} 
{\em Quantum Magnetism}, Lecture Notes in Physics Vol.~645, 
edited by U.~Schollw{\"{o}}ck, J.~Richter, D.~J.~J.~Farnell, and R.~F.~Bishop
(Springer-Verlag, Berlin, 2004).

\bibitem{2D_magnetism_2} 
G.~Misguich and C.~Lhuillier, 
in {\it Frustrated Spin Systems}, edited by H.~T.~Diep 
(World Scientific, Singapore, 2005), p.~229.

\bibitem{Sachdev:2011}
S.~Sachdev and B.~Keimer, 
Phys.\ Today {\bf 64}(2), 29 (2011).

\bibitem{Villain:1977}
J.~Villain,
J.\ Phys.\ (France) {\bf 38}, 385 (1977);
J.~Villain, R.~Bidaux, J.~P.~Carton, and R.~Conte,
{\it ibid.}\/ {\bf 41}, 1263 (1980).

\bibitem{Singh:1998}
R.~R.~P.~Singh, O.~A.~Starykh, and P.~J.~Freitas,
J.\ Appl.\ Phys.\ {\bf 83}, 7387 (1998).

\bibitem{Palmer:2001}
S.~E.~Palmer and J.~T.~Chalker,
Phys.\ Rev.\ B {\bf 64}, 094412 (2001).

\bibitem{Brenig:2002}
W.~Brenig and A.~Honecker,
Phys.\ Rev.\ B {\bf 65}, 140407(R) (2002).

\bibitem{Canals:2002}
B.~Canals,
Phys.\ Rev.\ B {\bf 65}, 184408 (2002).

\bibitem{Starykh:2002}
O.~A.~Starykh, R.~R.~P.~Singh, and G.~C.~Levine,
Phys.\ Rev.\ Lett.\ {\bf 88}, 167203 (2002).

\bibitem{Sindzingre:2002}
P.~Sindzingre, J.-B.~Fouet, and C.~Lhuillier,
Phys.\ Rev.\ B {\bf 66}, 174424 (2002).

\bibitem{Fouet:2003}
J.-B.~Fouet, M.~Mambrini, P.~Sindzingre, and C.~Lhuillier,
Phys.\ Rev.\ B {\bf 67}, 054411 (2003).

\bibitem{Berg:2003}
E.~Berg, E.~Altman, and A.~Auerbach,
Phys.\ Rev.\ Lett.\ {\bf 90}, 147204 (2003).

\bibitem{Tchernyshyov:2003}
O.~Tchernyshyov, O.~A.~Starykh, R.~Moessner, and A.~G.~Abanov,
Phys.\ Rev.\ B {\bf 68}, 144422 (2003).

\bibitem{Moessner:2004}
R.~Moessner, O.~Tchernyshyov, and S.~L.~Sondhi,
J.\ Stat.\ Phys.\ {\bf 116}, 755 (2004).

\bibitem{Hermele:2004}
M.~Hermele, M.~P.~A.~Fisher, and L.~Balents,
Phys.\ Rev.\ B {\bf 69}, 064404 (2004).

\bibitem{Brenig:2004}
W.~Brenig and M.~Grzeschik,
Phys.\ Rev.\ B {\bf 69}, 064420 (2004).

\bibitem{Bernier:2004}
J.~S.~Bernier, C.~H.~Chung, Y.~B.~Kim, and S.~Sachdev,
Phys.\ Rev.\ B {\bf 69}, 214427 (2004).

\bibitem{Starykh:2005}
O.~A.~Starykh, A.~Furusaki, and L.~Balents,
Phys.\ Rev.\ B {\bf 72}, 094416 (2005).

\bibitem{Schmidt:2006}
H.-J.~Schmidt, J.~Richter, and R.~Moessner,
J.\ Phys.\ A: Math.\ Gen.\ {\bf 39}, 10673 (2006).

\bibitem{Arlego:2007}
M.~Arlego and W.~Brenig,
Phys.\ Rev.\ B {\bf 75}, 024409 (2007); {\it ibid.}~{\bf 80}, 099902(E) (2009).

\bibitem{Moukouri:2008}
S.~Moukouri,
Phys.\ Rev.\ B {\bf 77}, 052408 (2008).

\bibitem{Chan:2011}
Y.-H.~Chan, Y.-J.~Han, and L.-M.~Duan,
Phys.\ Rev.\ B {\bf 84}, 224407 (2011).

\bibitem{Bishop:2012_checkerboard}
R.~F.~Bishop, P.~H.~Y.~Li, D.~J.~J.~Farnell, J.~Richter, and C.~E.~Campbell,
Phys.\ Rev.\ B {\bf 85}, 205122 (2012).

\bibitem{Emery:2000}
V.~J.~Emery, E.~Fradkin, S.~A.~Kivelson, and T.~C.~Lubensky,
Phys.\ Rev.\ Lett.\ {\bf 85}, 2160 (2000).

\bibitem{Mukhopadhyay:2001}
R.~Mukhopadhyay, C.~L.~Kane, and T.~C.~Lubensky,
Phys.\ Rev.\ B {\bf 64}, 045120 (2001).

\bibitem{Vishwanath:2001}
A.~Vishwanath and D.~Carpentier,
Phys.\ Rev.\ Lett.\ {\bf 86}, 676 (2001).

\bibitem{Bishop_1987:ccm}
R.~F.~Bishop and H.~G.~K\"{u}mmel,
Phys.\ Today {\bf 40}(3), 52 (1987).

\bibitem{Arponen:1991_ccm}
J.~S.~Arponen and R.~F.~Bishop,
Ann.\ Phys.\ (N.Y.) {\bf 207}, 171 (1991).

\bibitem{Bi:1991} 
R.~F.~Bishop, 
Theor.\ Chim.\ Acta {\bf 80}, 95 (1991).

\bibitem{Bishop:1998} 
R.~F.~Bishop,  
in {\em Microscopic Quantum Many-Body Theories and Their Applications}, 
Lecture Notes in Physics Vol. 510, edited by J.~Navarro and A.~Polls,
(Springer-Verlag, Berlin, 1998), p.~1.

\bibitem{Fa:2004} 
D.~J.~J.~Farnell and R.~F.~Bishop, 
in {\em Quantum Magnetism}, 
Lecture Notes in Physics Vol. 645,
edited by U.~Schollw{\"{o}}ck, J.~Richter, D.~J.~J.~Farnell, and R.~F.~Bishop, 
(Springer-Verlag, Berlin, 2004), p.~307.

\bibitem{Bishop:1991_ccm}
R.~F.~Bishop, J.~B.~Parkinson, and Yang Xian, 
Phys.\ Rev.\ B {\bf 44}, 9425 (1991).

\bibitem{Ze:1998} 
C.~Zeng, D.~J.~J.~Farnell, and R.~F.~Bishop, 
J.\ Stat.\ Phys. {\bf 90}, 327 (1998).

\bibitem{Kr:2000} 
S.~E.~Kr{\"{u}}ger, J.~Richter, J.~Schulenburg, D.~J.~J.~Farnell, and R.~F.~Bishop, 
Phys.\ Rev.\ B {\bf 61}, 14607 (2000).

\bibitem{Bishop:2000} 
R. F. Bishop, D. J. J. Farnell, S.E. Kr\"uger, J. B. Parkinson, J. Richter, and C. Zeng, 
J.\ Phys.: Condens.\ Matter {\bf 12}, 6887 (2000).

\bibitem{Fa:2001}
D.~J.~J.~Farnell, R.~F.~Bishop, and K.~A.~Gernoth,
Phys.\ Rev.\ B {\bf 63}, 220402(R) (2001).

\bibitem{Darradi:2005} 
R. Darradi, J. Richter, and D.~J.~J.~Farnell,
Phys.\ Rev.\ B {\bf 72}, 104425 (2005).
% Shastry Sutherland

\bibitem{Darradi:2008_J1J2mod} 
R.~Darradi, O.~Derzhko, R.~Zinke, J.~Schulenburg, S.~E.~Kr\"uger, and J.~Richter,
 %      Ground-state phases of the spin-$1/2$ $J_1$--$J_2$ Heisenberg
 %      antiferromagnet on the square lattice:
 %      A high-order coupled cluster treatment\\
Phys.\ Rev.\ B {\bf 78}, 214415 (2008).

\bibitem{schmalfuss}  
D. Schmalfu{\ss}, R. Darradi, J. Richter, J.~Schulenburg,  and D.~Ihle,
Phys.\ Rev.\ Lett.\ {\bf 97}, 157201 (2006).

\bibitem{Bi:2008_JPCM} 
R.~F.~Bishop, P.~H.~Y.~Li, R.~Darradi, and J.~Richter, 
J.\ Phys.: Condens.\ Matter {\bf 20}, 255251 (2008).

\bibitem{Bi:2008_PRB_J1xxz_J2xxz} 
R.~F.~Bishop, P.~H.~Y.~Li, R.~Darradi, J.~Schulenburg, and J.~Richter,
Phys.\ Rev.\ B {\bf 78}, 054412 (2008).

\bibitem{Bi:2009_SqTriangle}
R.~F.~Bishop, P.~H.~Y.~Li, D.~J.~J.~Farnell, and C.~E.~Campbell,
Phys.\ Rev.\ B {\bf 79}, 174405 (2009).

\bibitem{richter10}   
J. Richter, R. Darradi, J.~Schulenburg, D.J.J. Farnell, and H. Rosner,
%        Frustrated spin-$1/2$ $J_1$-$J_2$ Heisenberg ferromagnet
%        on the square lattice
%        studied via exact diagonalization and coupled-cluster method\\
Phys.\ Rev.\ B {\bf 81}, 174429 (2010).

\bibitem{UJack_ccm}
R.~F.~Bishop, P.~H.~Y.~Li, D.~J.~J.~Farnell, and C.~E.~Campbell,
Phys.\ Rev.\ B {\bf 82}, 024416 (2010).

\bibitem{Reuther:2011_J1J2J3mod}
J.~Reuther, P.~W\"{o}lfle, R.~Darradi, W.~Brenig, M.~Arlego, and J.~Richter,
Phys.\ Rev.\ B {\bf 83}, 064416 (2011).

\bibitem{Farnell:2011}
D.~J.~J.~Farnell, R.~F.~Bishop, P.~H.~Y.~Li, J.~Richter, and C.~E.~Campbell,
Phys.\ Rev.\ B {\bf 84}, 012403 (2011).

\bibitem{Gotze:2011}
O.~G{\"o}tze, D.~J.~J.~Farnell, R.~F.~Bishop, P.~H.~Y.~Li, and J.~Richter,
Phys.\ Rev.\ B {\bf 84}, 224428 (2011).


\bibitem{Li:2012_honey_full}
P.~H.~Y.~Li, R.~F.~Bishop, D.~J.~J.~Farnell, and C.~E.~Campbell,
Phys.\ Rev.\ B {\bf 86}, 144404 (2012).

\bibitem{Li:2012_anisotropic_kagomeSq}
P.~H.~Y.~Li, R.~F.~Bishop, C.~E.~Campbell, D.~J.~J.~Farnell, O.~G{\"o}tze and J.~Richter,
Phys.\ Rev.\ B {\bf 86}, 214403 (2012).

\bibitem{Struck:2011}
J.~Struck, C.~{\"O}lsch{\"a}ger, R.~Le Targat, P.~Soltan-Panahi, A.~Eckardt, 
M.~Lewenstein, P.~Windpassinger, and K.~Sengstock,
Science {\bf 333}, 996 (2011).

\bibitem{Bethe:1931}
H.~Bethe, 
Z.\ Phys. {\bf 71}, 205 (1931).

\bibitem{cccm} We use the program
package CCCM of D.~J.~J. Farnell and J. Schulenburg, see
http://www-e.uni-magdeburg.de/jschulen/ccm/index.html.

\bibitem{Zh:1991}
Zheng Weihong, J.~Oitmaa, and C.~J.~Hamer,
Phy.\ Rev.\ B {\bf 43}, 8321 (1991).

\bibitem{Sa:1997}
A.~W.~Sandvik,
Phys.\ Rev.\ B {\bf 56}, 11678 (1997).

\bibitem{Ma:1955}
W.~Marshall, 
Proc.\ R.\ Soc.\ London, Ser.\ A {\bf 232}, 48 (1955).

\bibitem{Zheng:2006}
W.~Zheng, J.~O.~Fjaerestad, R.~R.~P.~Singh, R.~H.~McKenzie, and R.~Coldea,
Phys.\ Rev.\ B {\bf 74}, 224420 (2006).

\bibitem{Ca:1999}
L.~Capriotti, A.~E.~Trumper, and S.~Sorella,
Phys.\ Rev.\ Lett. {\bf 82}, 3899 (1999).

\bibitem{Hulthen:1938}
L.~Hulth\'{e}n, Ark.\ Mat.\ Astron.\ Fys.\ A {\bf 26} (No.\ 11), 1 (1938);
R.~Orbach, Phy.\ Rev.\ {\bf 112}, 309 (1958); 
C.~N.~Yang and C.~P.~Yang, {\it ibid.} {\bf 150}, 321 (1966);
{\bf 150}, 327 (1966);
R.~J.~Baxter, J.\ Stat.\ Phys.\ {\bf 9}, 145 (1973).

\end{thebibliography}
\end{document}